\documentclass[12pt,subeqn,a4paper]{article}
\usepackage{amssymb,amsmath,amsfonts,amsthm, amscd, mathrsfs,helvet}
\usepackage{bm}
\usepackage{graphicx,verbatim}
\usepackage{psfrag}
\usepackage[all]{xy}

\makeatletter \@addtoreset{equation}{section} \makeatother
\makeatletter \@addtoreset{figure}{section} \makeatother

\def\CG{{\cal G}}

\def\CO{{\cal O}}

\def\CW{{\cal W}}
\def\CZ{{\cal Z}}

\def\g{\gamma}
\def\d{\delta}\def\e{\epsilon}
\def\th{\theta}
\def\l{\lambda}

\def\r{\rho}
\def\t{\tau}
\def\w{\omega}\def\G{\Gamma}
\def\D{\Delta}\def\L{\Lambda}
\def\O{\Omega}

\newcommand{\EQ}[1]{\begin{equation}\begin{split} #1
\end{split}\end{equation}}

\def\be{{\bar e}}

\def\r2{\sqrt{2}}

\def\bea{\begin{eqnarray} }
\def\eea{\end{eqnarray}}
\def\be{\begin{equation} }
\def\ee{\end{equation}}
\def\nn{\nonumber}

\begin{document}
\newcommand{\nd}[1]{/\hspace{-0.5em} #1}


\begin{titlepage}
\begin{flushright}
{\bf March 2011} \\
\end{flushright}
\begin{centering}
\vspace{.2in}
 {\large {\bf Quantization of Integrable Systems and a 2d/4d
     Duality }}

\vspace{.2in}

Nick Dorey and	Sungjay Lee \\
\vspace{.1 in}
DAMTP, Centre for Mathematical Sciences \\
University of Cambridge, Wilberforce Road \\
Cambridge CB3 0WA, UK \\
\vspace{.15in}
and  Timothy J. Hollowood\\
\vspace{.1 in}
Department of Physics\\
Swansea University\\
Swansea SA2 8PP, UK\\

%
%
\vspace{.2in}
{\bf Abstract} \\
\end{centering}
We present a new duality between the F-terms of supersymmetric
field theories defined in two- and
four-dimensions respectively. The duality relates
${\cal N}=2$ supersymmetric gauge theories in four dimensions, deformed
by an $\Omega$-background in one plane, to ${\cal N}=(2,2)$ gauged
linear $\sigma$-models in two dimensions.
On the four dimensional side,
our main example is ${\cal N}=2$
SQCD with gauge group $G=SU(L)$ and
$N_{F}=2L$ fundamental flavours. Using ideas of
Nekrasov and Shatashvili, we argue that the Coulomb branch of this
theory provides a quantization of the classical
Heisenberg $SL(2)$ spin chain. Agreement with the standard
quantization via the Algebraic Bethe Ansatz implies the
existence of an isomorphism between the chiral ring of the 4d theory and
that of a certain two-dimensional theory. The latter can be understood as
the worldvolume theory on a surface operator/vortex string probing the
Higgs branch of the same 4d theory. We check the proposed duality by
explicit calculation at low orders in the instanton expansion.
One striking consequence is that the Seiberg-Witten solution
of the 4d theory is captured by a one-loop computation in two
dimensions. The duality also has interesting connections with the AGT
conjecture, matrix models and topological
string theory where it corresponds to a refined version of the
geometric transition.

\end{titlepage}

\parskip 0.1 cm
\tableofcontents
\renewcommand{\thefootnote}{\#\arabic{footnote}}
\setcounter{footnote}{0}

\paragraph{}
\section{Introduction}
\paragraph{}
Supersymmetric gauge theories in two dimensions exhibit intriguing
similarities to their four dimensional counterparts which have been
noted many times in the past. Their common features include the
existence of protected quantities with holomorphic dependence on
F-term couplings and a related spectrum of BPS states which undergo
non-trivial monodromies and wall-crossing transitions in the space of
couplings/VEVs. In this paper we will propose a precise duality between
specific theories in four- and two-dimensions (henceforth denoted
Theory I and Theory II respectively). The
duality applies to the large class of four dimensional theories with
${\cal N}=2$ supersymmetry which can be realised by the standard
quiver construction as in \cite{WM}. As our main example we have,
\paragraph{}
{\bf Theory I}: Four-dimensional ${\cal N}=2$ SQCD with gauge
group $SU(L)$, $L$ hypermultiplets in the fundamental
representation with masses $\vec{m}_{F}=
(m_{1},\ldots,m_{L})$
and $L$ hypermultiplets in the anti-fundamental
with masses $\vec{{m}}_{AF}=(\tilde{m}_{1},\ldots,\tilde{m}_{L})$. 
The theory is conformally invariant in the UV with marginal coupling
$\tau=4\pi i/g^{2}\,+\,\vartheta/2\pi$.
\paragraph{}
For some purposes it will also be useful to consider the corresponding
$U(L)$ gauge theory. We consider Theory I in the presence of a particular
\footnote{As we explain in Section 2.2 below there are a
  family of inequivalent deformations related to each other by the
  low-energy electromagnetic duality group of the four-dimensional
  theory.} Nekrasov deformation with parameter $\epsilon$ which preserves
${\cal N}=(2,2)$ supersymmetry in an $\mathbb{R}^{1,1}$ subspace of
four-dimensional spacetime.  The resulting
effective theory in two dimensions is characterised by a (twisted)
superpotential, ${\mathcal W}^{(I)}$ with holomorphic dependence on
(twisted) chiral superfields. The superpotential
${\mathcal W}^{(I)}$ receives an infinite
series of corrections from perturbation theory and instantons which
encode the four-dimensional origin of the theory.
It has an $L$-dimensional lattice
of stationary points corresponding to supersymmetric vacua of the
deformed theory. These are determined by the F-term equation,
\bea
\vec{a}\,=\,\vec{m}_{F}\,-\,\vec{n}\epsilon & \qquad{} &
\vec{n}\,=\,(n_{1},\ldots,n_{L})\,\in\,\mathbb{Z}^{L} \nn
\eea
where $\vec{a}=(a_{1},\ldots,a_{L})$ are the usual special K\"{a}hler
coordinates on the Coulomb branch of the four-dimensional theory.
A generic point on the Coulomb branch of the undeformed
theory can be recovered in an appropriate $\epsilon\rightarrow 0$,
$|\vec{n}|\rightarrow \infty$ limit.
\paragraph{}
We will propose an exact duality
of Theory I to a surprisingly simple model defined in
two-dimensions which holds for all positive values
of the integers $\{n_{l}\}$ introduced above;
\paragraph{}
{\bf Theory II}: Two-dimensional ${\cal N}=(2,2)$ supersymmetric
Yang-Mills theory with gauge group $U(N)$ with $L$ chiral
multiplets in the fundamental representation with twisted masses
$\vec{M}_{F}=(M_{1},\ldots,M_{L})$ and $L$ chiral
multiplets in the anti-fundamental with twisted masses $\vec{{M}}_{{AF}}=
(\tilde{M}_{1},\ldots,\tilde{M}_{L})$. In addition the theory has a
single chiral multiplet in the adjoint representation with mass
$\epsilon$. The FI parameter $r$ and 2d vacuum angle $\theta$ combine to form
a complex marginal coupling $\hat{\tau}=ir+\theta/2\pi$.
\paragraph{}
Theory II has a twisted effective superpotential $\mathcal{W}^{(II)}$
which is one-loop exact \cite{Wphase}.
In both Theory I and Theory II, the
superpotential determines the chiral ring of
supersymmetric vacuum states.
\paragraph{}
{\bf Claim}: The chiral rings of Theory I and Theory II are
isomorphic. In particular, there is a
$1$-$1$ correspondence between
the supersymmetric vacua of the
two theories and, with an appropriate
identification of complex parameters, the values of the twisted
superpotentials coincide in corresponding vacua
(up to a vacuum-independent additive constant),
\bea
\begin{array}{c} \\ \mathcal{W}^{(I)} \end{array} &
\begin{array}{c} _{\rm{on-shell}} \\ {\equiv}
\end{array} & \begin{array}{c} \\ \mathcal{W}^{(II)} \end{array} \nn
\eea
The rank $N$ of the 2d gauge group is identified in terms of the 4d
parameters according to $N+L=\sum_{l=1}^{L}n_{l}$. Thus,
when $|\e|$ is small, low values of $N$
correspond to points near the Higgs branch root of
the 4d theory. The deformation parameter
$\epsilon$ of
Theory I is identified with adjoint mass of Theory II. The
explicit map between the remaining parameters takes the form,
\begin{align}
\hat{\tau}=\tau+\frac{1}{2}(N+1) \ , \qquad
\vec{M}_{F} = \vec{m}_{F} - \frac32 \vec{\e} \ ,
  \qquad
  \vec{{M}}_{{AF}} = \vec{{m}}_{{AF}} + \frac12 \vec{\e}\ .
\label{ident} \end{align}
where $\vec{\epsilon}=(\epsilon,\e,\ldots,\e)$. Further details of the
map between the chiral rings of the two theories is given in
Subection 2.5 below.
\paragraph{}
The initial motivation for this duality comes from the mysterious connection
between supersymmetric gauge theories and quantum integrable systems
developed in a remarkable series of papers by Nekrasov and Shatashvili (NS)
\cite{NS2d,NS4d}. These authors propose a general correspondence
in which the space of supersymmetric vacua of a theory
with ${\cal N}=(2,2)$ supersymmetry is identified with the Hilbert space of
a quantum integrable system. The generators of the chiral ring are
mapped to the commuting conserved charges of the integrable
system. The twisted superpotential itself corresponds to the so-called
Yang-Yang potential which is naturally thought of as a generating
function for the conserved charges.  The ideas of \cite{NS4d} also extend the
well known connection between ${\cal N}=2$ supersymmetric gauge theory
in four-dimensions and classical integrable systems
\cite{Gors,Mart,DW} which is reviewed in Section 2 below.
In particular they propose
that the introduction of a Nekrasov deformation in one plane,
breaking four-dimensional supersymmetry to an ${\cal N}=(2,2)$ subalgebra,
corresponds to a quantization of the corresponding
classical integrable system with the deformation parameter $\epsilon$
playing the role of Planck's constant $\hbar$.
\paragraph{}
Our main observation is that the same quantum integrable system
arises in two different contexts.
In the case $\epsilon=0$, it has been known for some time \cite{Marsh1} that
Theory I corresponds to a {\em classical\/} Heisenberg spin chain with spins
whose Poisson brackets provide a representation of
$\mathfrak{sl}(2)$ at each site. After imposing appropriate reality
conditions, we will adapt the ideas of \cite{NS4d} and argue that
introducing non-zero $\epsilon$ corresponds to a specific quantization
of this system\footnote{For other recent work relating ${\cal N}=2$
  SQCD and the quantum Heisenberg spin chain see \cite{Marsh,Zenk}}
in which the classical spins at each site
are replaced by quantum operators acting in a highest-weight
representation of $SL(2,\mathbb{R})$.
The resulting quantum chain is
integrable and can be diagonalised exactly using the Quantum Inverse
Scattering method which leads to a simple set of rational
Bethe Ansatz equations. However, precisely these equations also arise
as the F-term equations of Theory II and the corresponding
twisted superpotential, $\mathcal{W}^{(II)}$,
coincides with the Yang-Yang potential of the spin
chain \cite{NS2d}. This is not a coincidence:
with the identification of parameters proposed above,
Theory II can be identified as the
worldvolume theory of a vortex string or surface
operator probing the Higgs branch of Theory I\footnote{There are some
  subtleties to this relation which we discuss in Subsection 2.4
  below}. As we discuss below,
this connection suggests a physical explanation of the correspondence
along the lines of \cite{Sh,HT} as well as relations to several other
recent developments.
\paragraph{}
Equivalence between the NS quantization and the standard quantization
of the spin chain implies the duality between Theories I and II
proposed above. In Section 3 below we test the duality by
an explicit calculation of $\mathcal{W}^{(I)}$ in each vacuum 
including classical, perturbative contributions as well as
non-perturbative contributions up to
second order in the four-dimensional instanton expansion. The
corresponding computation of $\mathcal{W}^{(II)}$ involves an
iterative solution of the one-loop exact F-term equations in powers
of the parameter $q=\exp(2\pi i \tau)$.
The calculation yields precise agreement in all vacua when the
parameters are identified according to (\ref{ident}).
\paragraph{}
One interesting consequence of the proposed duality is that the
full Seiberg-Witten solution of Theory I can be recovered by taking an
$N\rightarrow\infty$, $\epsilon\rightarrow 0$ limit of the
one-loop exact F-term equations of Theory II. In fact this is just the
standard semiclassical limit of the non-compact spin chain (see {\it e.g.\/}
\cite{BGK1,BGK2}) where the
Bethe roots condense to form branch cuts in the spectral plane. In our
context the resulting double cover of the complex plane is the Seiberg-Witten
curve. This
is very reminiscent of the Dijkgraaf-Vafa matrix model \cite{DV} approach where
the eigenvalues of an $N\times N$ matrix
condense to form the branch cuts of the
Seiberg-Witten curve. An important
difference is that, in the NS limit, the Bethe Ansatz equations provide
an exact solution of the system even at finite $N$ (see point {\bf 4} below).
\paragraph{}
In a forthcoming paper \cite{us2}
we will also sketch a similar correspondence for a larger
class of ${\cal N}=2$ quiver theories in four (as well as five and
six) dimensions.
A common feature is that quantization leads to
spin chains based on highest weight
representations of Lie algebras where the
ferromagnetic ground-state of the spin chain corresponds to the root of
the gauge theory Higgs branch. In all cases, the Algebraic
Bethe Ansatz leads to simple
rational, (trigonometric, elliptic) equations which coincide with the
F-term equations of a dual two-dimensional theory.
In forthcoming work \cite{us2} we will show how the correspondence can be
proved by directly relating the Bethe Ansatz equation to the
saddle-point equation describing the instanton density in the
Nekrasov-Shatashvili limit
\cite{NS4d} (see also the recent papers \cite{Pog,Fucito:2011pn}).
\paragraph{}
There are also several interesting connections to other developments
in supersymmetric gauge theory both new and old:
\paragraph{}
{\bf 1:} As already mentioned,
there is a simple physical relation between the two
theories which holds with identifications between the parameters
described above:
Theory II can be understood as the theory on the world-volume
of a vortex string or surface operator
\cite{AGGTV,DGH} probing the Higgs branch of the same four-dimensional
theory. More precisely, Theory II is a gauged linear
$\sigma$-model for the moduli
space\footnote{There are some subtleties to this relation which are
  discussed in Section 2.4 below} 
of $N$ non-abelian vortices of Theory I \cite{HT1}
(See also \cite{Yoshida,Bonelli}).
The proposed duality relates the world-volume theory of $N$ vortices
to the bulk theory ({\it i.e.\/}~Theory I) on its Coulomb
branch. As usual, Higgs phase vortices carry quantized magnetic flux.
Interestingly, the dual Coulomb branch vacua of Theory I
also exhibit quantized magnetic fluxes in the presence of the
$\Omega$-deformation.
\paragraph{}
{\bf 2:} The duality provides an analog of
the geometric transition of Gopakumar and Vafa \cite{GV} for the
``refined'' topological string \cite{ref} with refinement parameters
$\epsilon_{1}\neq \epsilon_{2}$ in the Nekrasov-Shatashvili (NS) limit
$\epsilon_{2}\rightarrow 0$. Theories I and II
correspond to the closed and open string sides of the transition
respectively.
\paragraph{}
{\bf 3:} The duality can also be understood in terms of the conjecture
\cite{AGT} relating four dimensional supersymmetric gauge theory with
Liouville theory. We will argue that in certain cases the proposed
duality is equivalent to the conjecture of Alday {\it et al} \cite{AGGTV}
that a surface operator in gauge theory corresponds to a particular
degenerate operator in Liouville theory.
\paragraph{}
{\bf 4:} Recent work \cite{DV1} has advocated a further duality
between ${\cal N}=2$ supersymmetric gauge theories in a four
dimensional $\Omega$-background
and matrix models. In particular, the Nekrasov partition function of
Theory I should have a matrix integral representation. Here we identify the
dimension of matrix as the rank $N$ of the gauge group of Theory II.
We conjecture that the resulting integral over the matrix
eigenvalues can be evaluated explicitly by a saddle point in the NS
limit {\em even at finite $N$}. The saddle-point equations are
precisely the Bethe Ansatz equations of the spin chain and the free
energy is equal to the prepotential of the gauge theory.
This should also have interesting consequences for refined open topological
string amplitudes in the limit $\epsilon_{2}\rightarrow 0$.
\paragraph{}
{\bf 5:} Some time ago a duality was proposed \cite{D1,DHT}
relating the BPS spectrum of a two-dimensional theory and that of
a corresponding four-dimensional gauge theory at the root of
its Higgs branch. The correspondence is further
studied in \cite{Lee:2009fc} with care given to precise supermultiplet
counting together with their wall-crossing behaviour.
We show that the present proposal reduces to the earlier one in the limit
$\epsilon\rightarrow 0$.
\paragraph{}
The rest of the paper is organised as follows. In Section 2, we review
the basic features of Theory I, its relation to the classical
Heisenberg spin chain in the case $\epsilon=0$ and its quantization
for $\epsilon\neq 0$. We  also introduce Theory II and review
its realisation on the worldvolume of a vortex string/surface
operator and provide the precise statement of our duality
conjecture. Section 3 is devoted to a detailed check of this
proposal. Discussion of our results, generalisations and connections to
topological strings, matrix models and
Liouville theory are presented in Section 4.

\section{Supersymmetric Gauge Theory and Integrable Systems}
\paragraph{}
In this Section we will begin reviewing the relevant feature of the
four-dimensional theory introduced in Section 1. As above, we focus
on four-dimensional ${\cal N}=2$ Super QCD with gauge group $G=SU(L)$ and
$N_{F}=2L$ hypermultiplets. For an $SU(L)$ gauge theory,
hypermultiplets in the fundamental and
anti-fundamental representations of the gauge group
are essentially equivalent. It is nevertheless
convenient to focus on the case where
half of the $2L$ hypermultiplets
are in the fundamental representation
and the rest in the anti-fundamental representation. The duality
discussed below will apply equally to the corresponding $U(L)$ 
gauge theory which differs from the $SU(L)$ theory by an additional
$U(1)$ factor which is IR free. 
The fundamental and anti-fundamental
hypermultiplet masses are denoted by $m_{l}$ and
$\tilde{m}_{l}$ with $l=1,2,\ldots, L$ respectively.
The $SU(L)$ theory is conformally
invariant in the UV with marginal coupling
$\tau=4\pi i/g^{2}\,+\,\vartheta/2\pi$.
\paragraph{}
The exact
low-energy solution of the undeformed theory is governed by the
corresponding Seiberg-Witten curve,
\begin{align}
  \prod_{l=1}^{L} (v - \tilde m_l) t^2 - 2 \prod_{l=1}^{L}
  (v-\phi_l) t - h ( h+2) \prod_{l=1}^{L} (v -m_l ) = 0 \ .
  \qquad h(\tau)= -\frac{2q}{q+1} \ , \label{sw1}
\end{align}
where $q=\exp(2\pi i\tau)$ is the factor associated with a
four-dimensional Yang-Mills instanton.
Here $\phi_{l}$ denote the $L$ classical eigenvalues of the
adjoint scalar field $\varphi$ in the $\mathcal{N}=2$ vector multiplet.
As the gauge group is $SU(L)$ we impose a traceless condition
$\sum_{l=1}^{L}\phi_{l}=0$.
\paragraph{}
The masses of BPS states in the four-dimensional theory are determined
by a meromorphic differential, $\lambda_{\rm SW}=v
dt/t$ on the curve. A standard basis of A- and B-cycles on the
Seiberg-Witten curve, with $A_{l}\cap B_{m}=\delta_{lm}$
can be defined in the weak-coupling limit $\tau\rightarrow
i\infty$. Electric and magnetic central charges are determined by the
periods of $\lambda_{\rm SW}$ in this basis,
\bea \vec{a} \,\,=\,\, \frac{1}{2\pi i} \oint_{\vec{A}}\, \lambda_{\rm
  SW}\,\, , & &
\vec{a}^{D} \,\,=\,\,\frac{1}{2\pi i}\, \frac{\partial
  \mathcal{F}}{\partial \vec{a}}
 \,\,=\,\,
\frac{1}{2\pi i} \oint_{\vec{B}}\, \lambda_{\rm
  SW}\,\ ,   \label{SWl} \eea
where $\vec{a}=(a_{1},\ldots,a_{L})$ with similar notation for other
$L$-component vectors. For theories with matter the Seiberg-Witten
differential also has simple poles at the points $x=m_{l}$,
$\tilde{m}_{l}$ with residues $m_{l}$ and $\tilde{m}_{l}$
respectively. It is convenient to introduce additional cycles
$\vec{C}_{F}=({C}_{1},\dots,{C}_{L})$
and $\vec{C}_{AF}=(\tilde{C}_{1},\dots,\tilde{C}_{L})$
encircling these poles so that the
corresponding periods of $\lambda_{\rm SW}$ are,
\bea \vec{m}_{F} \,\,=\,\, \frac{1}{2\pi i} \oint_{\vec{C}_{F}}\, \lambda_{\rm
  SW}\,\, , & &
\vec{{m}}_{AF} \,\,=\,\,
\frac{1}{2\pi i} \oint_{\vec{{C}}_{AF}}\, \lambda_{\rm
  SW}\,\ .  \nn \eea
\begin{figure}
\centering
\includegraphics[width=130mm]{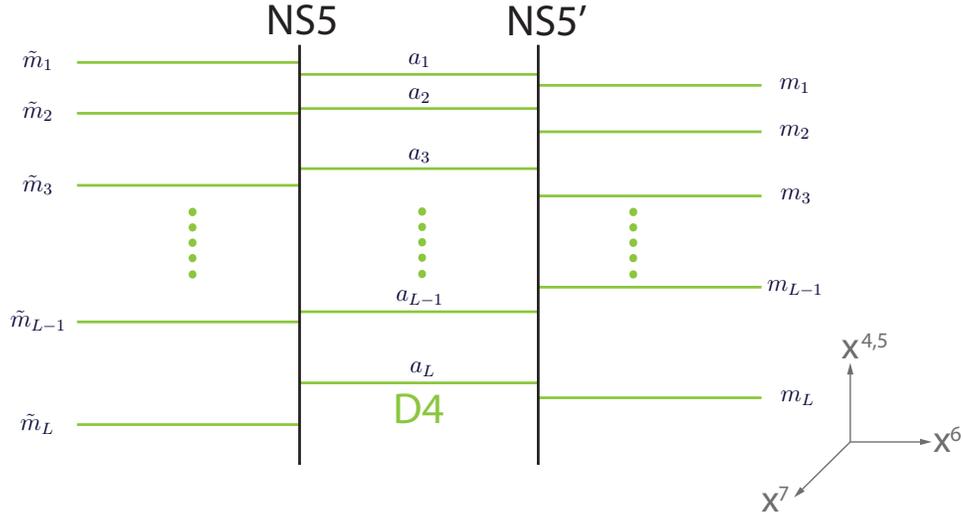}
\caption{IIA brane construction for a generic point on the Coulomb
branch of Theory I.}
\label{Qfig1}
\end{figure}
\begin{figure}
\centering
\includegraphics[width=130mm]{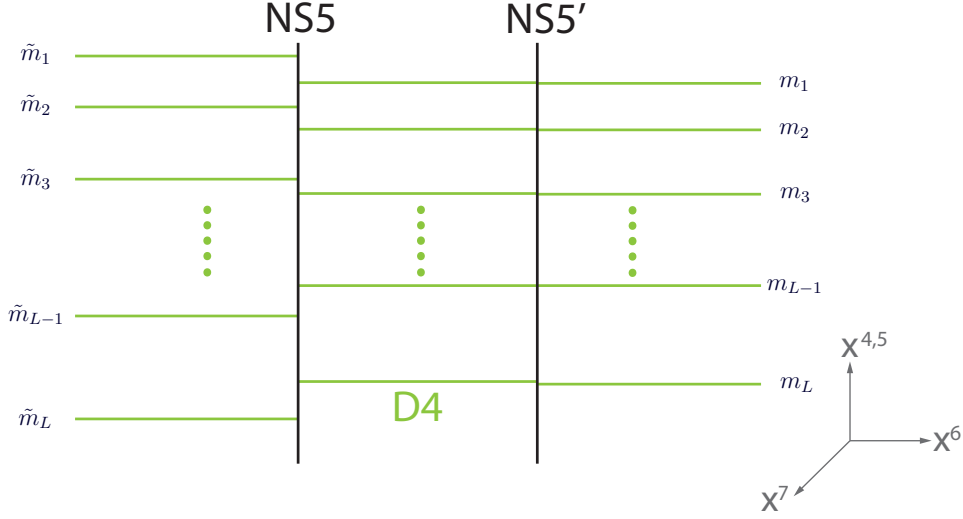}
\caption{The Higgs branch root $\vec{a}=\vec{m}_F$.}
\label{Qfig2}
\end{figure}

The standard IIA brane construction of Theory I at a generic point on
its Coulomb branch is shown in Figure
(\ref{Qfig1}). Here we follow the conventions of \cite{WM}.
Each horizontal line corresponds to a D4 brane. In the
figure each D4 is labelled by the corresponding value of the complex
coordinate $v=x_{4}+ix_{5}$. In the following, the point on the Coulomb
branch where it touches the Higgs branch will have a particular
significance. The configuration corresponding to this 
Higgs branch root is shown in Figure (\ref{Qfig2}). 
At the root of baryonic Higgs branch, the $\vec A - \vec C$ cycles
degenerate leading to $L-1$ additional massless hypermultiplets.
This corresponds to a factorization of the Seiberg-Witten curve. More
precisely, when $\phi_l$'s are tuned to satisfy a relation
\begin{align}\label{RH}
  - h \prod_{l=1}^L \big( v - \tilde m_l \big) + ( h+ 2) \prod_{l=1}^L
  \big( v - m_l \big) = 2 \prod_{l=1}^L \big(v - \phi_l \big)\ ,
\end{align}
the Seiberg-Witten curve becomes degenerate
\begin{align}\label{RHroot}
  \Big[ \prod_{l=1}^L \big(  v- \tilde m_l \big) t - (h+2)
\prod_{l=1}^L \big( v- m_l \big)
  \Big] \times \Big[ t + h \Big] = 0 \ ,
\end{align}
and $\vec A=\vec C_{F}$. 
We will soon explain a correspondence between the root of baryonic Higgs 
branch and ferromagnetic vacuum of the $SL(2,\mathbb{R})$ integrable 
model. 

\subsection{The classical integrable system}
\paragraph{}
We now review the connection between ${\cal N}=2$ supersymmetric gauge
theories in four dimensions and complex classical integrable systems. We begin
by introducing the Heisenberg spin chain.
\paragraph{}
We will consider a chain of $L$ complex ``spins'' \cite{BGK1,BBGK,BGK2}
corresponding to classical variables,
$\mathcal{L}^{\pm}_{l}$, $\mathcal{L}^{0}_{l}$, for $l=1,2,\ldots, L$
with Poisson brackets:
\bea
\{ \mathcal{L}^{+}_{l}, \mathcal{L}^{-}_{m}\} \,=\,
2i\delta_{lm} \mathcal{L}^{0}_{m} & \qquad{} &
\{ \mathcal{L}^{0}_{l}, \mathcal{L}^{\pm}_{m}\}
\,=\, \pm i\delta_{lm}
\mathcal{L}^{\pm}_{m}\ . \label{pb} \eea
Here $+$, $-$ and $0$ are indices in the Lie algebra
$\mathfrak{sl}(2)$. The spins at each site have a fixed value of the
quadratic Casimir,
\bea
\mathcal{L}^{+}_{l}\,\mathcal{L}^{-}_{l}\,\,+\,\, \left(
\mathcal{L}^{0}_{l}\right)^{2} & = & J^{2}_{l}\ . \label{Cas} \eea
\paragraph{}
Integrability of the classical spin chain starts from an auxiliary
linear problem based on the Lax matrix,
\bea
\mathbb{L}_{l}(x) & = & \left(\begin{array}{cc} x+
i\mathcal{L}^{0}_{l} & i\mathcal{L}^{+}_{l} \\ i\mathcal{L}^{-}_{l}
& x-i\mathcal{L}^{0}_{l} \end{array}\right)\ ,
\label{lax} \eea
where $x\in \mathbb{C}$ is a spectral parameter.
A tower of commuting conserved quantities are obtained by constructing the
corresponding monodromy matrix,
\bea
\mathbb{T}(x) & = & \mathbb{V}\,\prod_{l=1}^{L} \mathbb{L}_{l}(x-\theta_{l})\ ,
\nonumber
\eea
where we have included inhomogeneities $\theta_{l}$,
$l=1,2,\ldots, L$, at each site and a diagonal ``twist'' matrix,
\bea
\mathbb{V} & = & \left(\begin{array}{cc} -h & 0 \\ 0 & h+2
  \end{array}\right)\ . \nn \eea
As usual, the trace of the monodromy matrix is the generating function
for a tower of conserved charges,
\bea
2P(x) & = & {\rm tr}_{2}\left[ \mathbb{T}(x) \right] \nonumber \\
& =& 2x^{L}\,+\, q_{1}x^{L-1} \,+
\,\ldots\,+\,q_{L-1}x\, +\, q_{L}\ . \label{monod3} \eea
One may check starting from the Poisson brackets (\ref{pb}) that the
conserved charges, $q_{l}$, $l=1,2,\ldots L$ are in involution:
$\{q_{l}, \,\, q_{m}\}=0$, $\forall\, l,\,m$, which establishes the
Liouville integrability of the chain. The lowest charge $q_{1}$ can be
set to zero by a linear shift of the spectral parameter and we will do
so in the following.
\paragraph{}
As for any integrable Hamiltonian system, the exact classical
trajectories of the spins can be found by a canonical transformation
to action-angle variables. Using standard methods,
the action variables are identified as the moduli of a spectral curve
$\Gamma_{L}\subset \mathbb{C}^{2}$ defined by the equation,
\bea
F(x,y) & = & {\rm det}\left(y{\bf 1}_{2} \,\, - \,\, \mathbb{T}(x)\right)
\,\,=\,\,0\ ,
\nn
\eea
while the angle variables naturally parameterise the
Jacobian variety $\mathcal{J}(\Gamma_{L})$
More explicitly the curve takes the form,
\bea
\Gamma_{L}:\qquad{} y^{2}\,-\, 2\,P(x)\,y \,-\, h(h+2)\, K(x)\,
\tilde{K}(x) &=& 0\ ,
\label{sw2}
\eea
where,
\bea
K(x) \,=\, \prod_{l=1}^{L} \left(x-\theta_{l} - iJ_{l}\right) & \qquad{} &
\tilde{K}(x) \,=\, \prod_{l=1}^{L} \left(x-\theta_{l} + iJ_{l}\right)\ . \nn
\eea
This curve is an equivalent form of the Seiberg-Witten
curve (\ref{sw1}) for Theory I provided
we make the identifications,
\bea
\vec{m}_{F}=\vec{\theta} \,+\,i\vec{J} & \qquad{} & \vec{m}_{AF}=
\vec{\theta}\, -\, i\vec{J}\ ,
\nn
\eea
and
\bea
2P(x) & = & 2x^{L}\,+\, q_{2}x^{L-2} \,+
\,\ldots\,+\,q_{L-1}x\, +\, q_{L}  \nn \\
& = & 2\, \prod_{l=1}^{L}\, (x-\phi_{l}). \nn \eea
Also $h(\tau)$ is identified with the twist parameter of the spin
chain denoted by the same letter. The relation between (\ref{sw1}) and
(\ref{sw2}) corresponds to the holomorphic change of variables,
$t=y/\tilde{K}(x)$, $v=x$.
\paragraph{}
With this identification different values of the moduli of Theory I
are associated with different values of the integrals of motion
of the complex spin chain. One point of particular interest is the
ferromagnetic vacuum of the chain where each spin is in its classical
groundstate: $\vec{\mathcal{L}}^{0}_{l}=\vec{J}$,
$\vec{\mathcal{L}}^{\pm}=0$. It is easy to check that this corresponds to
the root of the Higgs branch in the gauge theory where the VEVs take
the values $\vec{a}=\vec{m}_{F}$.

\subsection{Nekrasov-Shatashvili Quantization}
\paragraph{}
We now turn our attention to the gauge theory in the presence of the so-called
$\Omega$-background. This deformation, which breaks four-dimensional
Lorentz invariance, is specified by parameters
$\epsilon_{1}$ and $\epsilon_{2}$. The ${\cal N}=2$ F-terms of the
deformed theory are determined by the Nekrasov partition
function \cite{N1,NO},
\bea
& \mathcal{Z}(\vec{a}, \epsilon_{1}, \epsilon_{2})\ . & \nn \eea
\paragraph{}
In this paper we will be mainly concerned with the
Nekrasov-Shatashvili limit $\epsilon_{2}\rightarrow 0$ with
$\epsilon_{1}$ held fixed, where the
deformation is restricted to one plane in $\mathbb{R}^{4}$. We define
a quantum prepotential in this limit as,
\bea
\mathcal{F}\left(\vec{a}, \epsilon\right) & = & \lim_{\epsilon_{2}
\rightarrow 0}\,
\left[ \left. \epsilon_{1}\epsilon_{2}\, \log\,
\mathcal{Z}(\vec{a}, \epsilon_{1},
\epsilon_{2})\right|_{\epsilon_{1}=\epsilon}\right]\ .
\nn \eea
In the further
limit $\epsilon\rightarrow 0$, the quantum prepotential reduces to
the familiar prepotential of the undeformed theory:
$\mathcal{F}(\vec a,\epsilon)\rightarrow \mathcal{F}(\vec a)$.
Following \cite{Marsh,Pop}, the quantum prepotential can
be obtained by a suitable deformation of the Seiberg-Witten
differential appearing in (\ref{SWl}),
\bea
\lambda(\epsilon) & = & \lambda_{\rm SW} \,\,+\,\,{\cal O}(\epsilon)\ ,
\nn \eea
with periods,
\bea \vec{a}(\epsilon) \,\,=\,\, \frac{1}{2\pi i} \oint_{\vec{A}}\,
\lambda(\epsilon) \,\, , & &
\vec{a}^{D}(\epsilon) \,\,=\,\,
\frac{1}{2\pi i} \oint_{\vec{B}}\, \lambda(\epsilon)\,\ ,   \label{aeps} \eea
such that,
\bea
\vec{a}^{D}(\epsilon)  & = & \frac{1}{2\pi i}\, \frac{\partial
  }{\partial \vec{a}(\epsilon)}\,\mathcal{F}(\vec{a},\epsilon)\ .
  \label{ade}
\eea
For convenience we will suppress the $\epsilon$ dependence of the
deformed central charges from now on and denote them simply as
$\vec{a}$ and $\vec{a}^{D}$.
\paragraph{}
For $\epsilon\neq 0$, the four-dimensional ${\cal N}=2$ supersymmetry
is broken down to ${\cal N}=(2,2)$ supersymmetry two dimensions.
The zero modes of the $U(1)^{L-1}$ vector multiplet in the
four-dimensional low energy theory give rise to a field strength
multiplet in two dimensions. This multiplet includes
the gauge field strength $\vec{F}_{01}$ in the undeformed
directions and the scalar
fields $\vec{a}$ which parametrize the 4d Coulomb branch.
Thus $\vec{a}$ is the lowest component of a twisted chiral
superfield in $\mathcal{N}=(2,2)$ superspace. This superfield inherits
a twisted superpotential from the partition function of the four-dimensional
theory. The resulting twisted superpotential is a multi-valued function on
the Coulomb branch,
\bea
\mathcal{W}^{(I)}\left(\vec{a},\epsilon\right) & = & \frac{1}{\epsilon}\,
\mathcal{F}\left(\vec{a},\epsilon\right) \,\,-\,\,2\pi i\,
\vec{k}\cdot
\vec{a}
\label{wa} \eea
where the integer-valued vector\footnote{For an $SU(L)$ theory we should
also impose $\sum_{l}k_{l}=0$.} $\vec{k}\in \mathbb{Z}^{L}$
corresponds to the choice of branch. This choice corresponds to the
freedom to shift the 2d vacuum angle associated with each $U(1)$
factor in the low energy gauge group by an integer multiple of
$2\pi$. This is equivalent to introducing a constant electric field
$\vec{F}_{01}$ in spacetime which is then screened 
by pair creation \cite{Col,Wphase}. The vector
$\vec{k}$ thus specifies the choice of a quantized two-dimensional
electric flux in the Cartan subalgebra of $SU(L)$.
\paragraph{}
The main claim of \cite{NS4d} is that $\mathcal{W}^{(I)}(\vec{a},\epsilon)$
is the Yang-Yang potential for a quantization of the classical
integrable system described above, in which the effective
Planck constant $\hbar$
is proportional to the deformation parameter $\epsilon$. This requires
further explanation: corresponding to a complex classical integrable
system there can be several choices of reality condition which yield
inequivalent real integrable systems. Each of these real
systems gives rise upon quantization to different quantum integrable
systems. As discussed in \cite{NS4d}, the twisted superpotential given
above gives rise to a particular quantization which they refer
to as Type A which we now review.
\paragraph{}
The F-term equations coming from (\ref{wa}) can written in terms of
the deformed magnetic central charge using (\ref{ade}) as follows,
\bea
\frac{\partial \mathcal{W}^{(I)}}{\partial \vec{a}}\,\,=\,\,0 & \Rightarrow &
\vec{a}^{D}\,\,\,\in\,\, \epsilon\,\mathbb{Z}^{L} \label{q1} \eea
This corresponds to a quantization condition for the conserved charges of
the integrable system, each point of the lattice $\mathbb{Z}^{L}$
corresponds to a different quantum state. The values of the commuting
conserved charges in each state are encoded in the on-shell value of the
superpotential $\mathcal{W}^{(I)}$.
Each branch of this multi-valued function corresponds to a supersymmetric
vacuum and to a state of the quantum integrable
system
Differentiating
with respect to the parameters yields the vacuum
expectation values of chiral operators in the corresponding
vacuum. For example,
\bea \langle {\rm Tr}\, \varphi^{2} \rangle & = & -\frac{\epsilon}{\pi i}
\frac{\partial}{\partial\tau}\, \left.\mathcal{W}^{(I)}\right|_{
\vec{a}^{D}\in\, \epsilon\mathbb{Z}^{L}}\ .  \nn \eea
To extract the VEVs of higher dimension chiral operators $\hat{O}_{k}
={\rm Tr}\,\varphi^{k}$ for $k>2$ we should deform the prepotential of
the UV theory with appropriate source terms \cite{Ext}.
 \paragraph{}
Theories with ${\cal N}=2$ supersymmetry in four dimensions exhibit
two distinct manifestations of electromagnetic duality. First there is
the low-energy
electromagnetic duality which provides an
alternative description of the low-energy effective theory at any point on
the Coulomb branch in terms of a dual field $\vec{a}^{D}$ with
dual prepotential ${\cal F}_{D}(\vec{a}_{D},\epsilon)$, which is
related to the original prepotential by Legendre
transform\footnote{Although the electro-magnetic duality
  transformation is modified and becomes a kind of Fourier transform
\cite{NO} for general deformation parameters
  $\epsilon_{1}$, $\epsilon_{2}$,
the standard transformation properties are regained in the NS limit.} .
\bea
\mathcal{F}_{D}\left(\vec{a}^{D}\right) & = &
\mathcal{F}\left(\vec{a}\right)\,\,- \,\,2\pi i
\,\vec{a}\cdot\vec{a}^{D}\ . \label{leg} \eea
In an $SU(L)$ gauge theory, the
full group of low-energy
duality transformation includes a copy of $Sp(2L,\mathbb{Z})$
which acts linearly on the central charges $(\vec{a},\vec{a}^{D})$.
For Theory I, there are also additional additive transformations
involving shifts of
the central charges by integer multiples of the mass
parameters $\vec{m}_{A}$, $\vec{{m}}_{AF}$ \cite{SW2}.
\paragraph{}
In \cite{NS4d}, the
authors proposed that these different formulations of the low-energy
theory give rise upon deformation to non-zero $\epsilon$,
to different quantizations of the same complex
classical integrable system. In particular, performing the
basic $\mathbb{Z}_{2}$ electric-magnetic duality
transformation we obtain a dual superpotential,
\bea
\mathcal{W}_{D}^{(I)}\left(\vec{a}_{D}\right) & = &
\mathcal{W}^{(I)}\left(\vec{a}\right)\,\, - \,\,\frac{2\pi i}{\epsilon}
\,\vec{a}\cdot\vec{a}^{D} \label{leg2} \eea
whose F-term equations give rise to dual quantization conditions,
denoted Type B in \cite{NS4d},
\bea
\frac{\partial \mathcal{W}^{(I)}_{D}}{\partial
\vec{a}^{D}}\,\,=\,\,0 & \Rightarrow &
\vec{a}\,\,\in\,\,\epsilon\,\mathbb{Z}^{L}. \label{q2} \eea
The discrete choice of vacuum now corresponds to a choice of magnetic
flux $\vec{F}_{23}$ in the Cartan subalgebra of the gauge group.
As a consequence of (\ref{leg2}), we note that the two superpotentials
$\mathcal{W}^{(I)}$ and
$\mathcal{W}^{(I)}_{D}$ are actually equal as multi-valued functions
when evaluated on-shell.  This is true for both 
quantization conditions (\ref{q1}) and (\ref{q2}).
\paragraph{}
Summarising the above,
the Type A and B quantization conditions can be written as,
 \bea \frac{1}{2\pi i} \oint_{\vec{A}}\,
\lambda(\epsilon) \,\, \in\,\, \epsilon\, \mathbb{Z}^{L}\, & {\rm and} &
\frac{1}{2\pi i} \oint_{\vec{B}}\, \lambda(\epsilon)\,\,\,\in\,\,
\epsilon\, \mathbb{Z}^{L}\ ,  \nn \eea
 respectively. Other quantizations corresponding
to other transformations in the duality group naturally correspond to
period conditions for different choices of basis cycles.
\paragraph{}
The ${\cal N}=2$ theory with gauge group $SU(L)$ and $2L$
hypermultiplets exhibits another form of electric-magnetic duality. The exact
S-duality of the theory relates electric and magnetic
observables of the theory at different
values of the marginal coupling $\tau$. In the present context it
implies a non-trivial duality between Type A and Type B
quantization at different values of $\tau$.

\subsection{Quantization via the Bethe Ansatz}
\begin{figure}
\centering
\includegraphics[width=130mm]{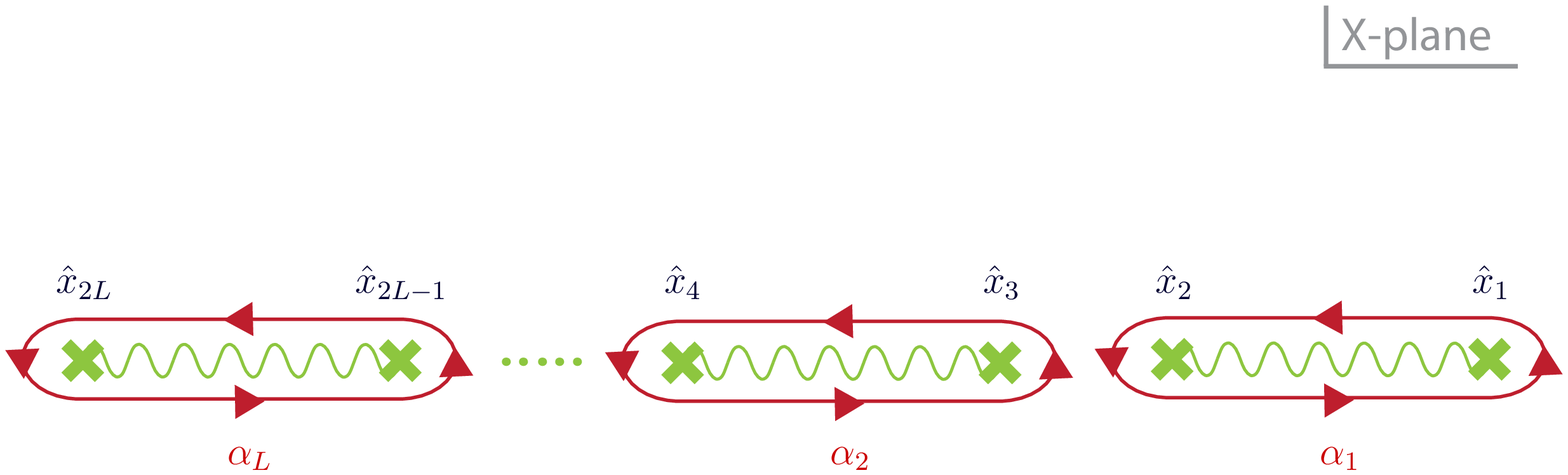}
\caption{The cut $x$-plane corresponding to the curve $\Gamma_{L}$.}
\label{Qfig}
\end{figure}

\paragraph{}
In this section we will review the standard approach to quantizing the
Heisenberg spin chain (see {\it e.g.\/}~\cite{Faddeev,Derk,BGK1,BGK2}).
Starting from the Poisson brackets (\ref{pb})
we make the usual
replacement of the classical variables
$\mathcal{L}^{\pm}_{l}$, $\mathcal{L}^{0}_{l}$ at each site
by operators $\hat{L}^{\pm}_{l}$, $\hat{L}^{0}_{l}$
obeying commutation relations,
\bea
[\hat{L}^{+}_{l}, \hat{L}^{-}_{m}] \,=\,
-2\,\hbar\, \delta_{lm} \hat{L}^{0}_{m} & \qquad{} &
[ \hat{L}^{0}_{l}, \hat{L}^{\pm}_{m}]
\,=\, \mp\, \hbar\, \delta_{lm}
\hat{L}^{\pm}_{m}\ , \label{qb} \eea
for $l,m=1,2,\ldots,L$. The spins at each site each commute with the
Casimir operator,
\bea
\hat{L}^{2}_{l} & = & \frac{1}{2}
\left(\hat{L}^{+}_{l}\hat{L}^{-}_{l}\,+\,
\hat{L}^{-}_{l}\hat{L}^{+}_{l}\right)\,\,+\,\,
\left(\hat{L}_{l}^{0}\right)^{2} \,\,\,
= \,\,\, s_{l}(s_{l}-1)\,\hbar^{2}.  \nn \eea
\paragraph{}
Depending on the value of $s_{l}$, these operators can
act on representations of either $SU(2)$ or $SL(2,\mathbb{R})$. We
will focus on the latter case. If we choose $s\in \mathbb{R}^{+}$
the spins can be chosen to
act in the principal discrete series representation\footnote{More
  precisely, when $s$
  is not equal to a half-integer, these are representations of the
  universal cover of $SL(2,\mathbb{R})$}
 of
$SL(2,\mathbb{R})$,
\bea
\mathcal{D}^{+}_{s} & = & \left\{ |s,\mu\rangle, \,\,\, \mu=s,\,
s+1,\,s+2,\ldots \right\}
\nn \eea
These are highest weight representations of $SL(2,\mathbb{R})$ and the
resulting spin chain admits a tower of commuting charges which can be
simultaneously diagonalised.
By restricting to a representation $\mathcal{D}^{+}_{s}$ with $s=s_{l}>0$
at the $l$'th site we are defining a {\em real} quantum integrable
system. An important feature of this non-compact spin chain is that it
has a semiclassical limit where each spin is highly excited.
As we discuss below, the relation to the {\em complex} integrable system we
discussed in subsection 2.1 becomes clear in this limit.
\paragraph{}
As we are dealing with a spin chain based on highest weight
representations the Algebraic Bethe Ansatz \cite{Faddeev}
is applicable and can be
used to find the exact spectrum of the model \cite{Derk}.
We now review the solution of the spin chain using an alternative
approach based on the Baxter equation.
We start by defining a quantum version of the classical Lax matrix
(\ref{lax}) which takes the form,
\bea
\hat{\mathbb{L}}_{l}(x) & = & \left(\begin{array}{cc} x+
i\hat{L}^{0}_{l} & i\hat{L}^{+}_{l} \\ i\hat{L}^{-}_{l}
& x-i\hat{L}^{0}_{l} \end{array}\right)
\label{qlax} \eea
and we define the quantum transfer matrix for an inhomogeneous chain
with twisted boundary conditions by,
\bea
\hat{T}(x) & = & {\rm
  tr}_{2}\left[\mathbb{V}\,\prod_{l=1}^{{\rightarrow}}\,
\hat{\mathbb{L}}_{l}(x-\theta_{l}) \right] \nn \\
& = & 2x^{L}\,+\, \hat{q}_{2}x^{L-2} \,+
\,\ldots\,+\,\hat{q}_{L-1}x\, +\, \hat{q}_{L}\ , \nn \eea
where $\{\hat{q}_{l}\}$, $l=1,2,\ldots,L$, are a set of mutually commuting
conserved charges which are the quantized versions of the
corresponding classical charges $\{q_{l}\}$.
\paragraph{}
The standard problem for a quantum integrable system is to find the
eigenstates of the transfer matrix,
\bea
 \hat{T}(x)\, \left|\Psi \right\rangle & = &t(x)\,
\left|\Psi \right\rangle\ , \nn \eea
where the eigenvalue, $t(x)$, is a polynomial of degree $L$ in the
spectral parameter $x$ by construction.
This is accomplished by allowing the transfer
matrix to act on a ``wavefunction'' $Q(x)$ which leads to the Baxter
equation,
\bea
-h a(x)\, Q(x+i\hbar)\,\,+\,\,(h+2)d(x) \,  Q(x-i\hbar) & = &
t(x)\, Q(x)\ , \label{baxter} \eea
where,
\bea
a(x) \,=\, \prod_{l=1}^{L} \left(x-\theta_{l} + is_{l}\hbar\right) &
\qquad{} &
d(x) \,=\, \prod_{l=1}^{L} \left(x-\theta_{l} - is_{l}\hbar \right) \nn
\eea
\paragraph{}
Looking for
solutions where $Q(x)$ is a polynomial of degree $N$,
\bea Q(x) & = & \prod_{j=1}^{N}\, \left(x-x_{j}\right) \nn \eea
we impose the polynomiality of $t(x)$ to obtain $N$ equations for the
$N$ zeros $x_{j}$ of $Q(x)$,
\bea
\prod_{l=1}^{L} \left(\frac{x_{j} - \theta_{l} - is_{l}\hbar}
{x_{j} - \theta_{l} + is_{l}\hbar}\right) & = &
{q}\,\,
\prod_{k\neq j}^{N}\, \left(\frac{x_{j}-x_{k} +i\hbar}
{x_{j}-x_{k} -i\hbar}\right)\ , \label{bae}
\eea
for $j=1,2,\ldots,N$ where ${q}=-h/(h+2)$.
These are the Bethe Ansatz Equations (BAE).
These equations are very well studied in the case of
an untwisted homogeneous $SL(2,\mathbb{R})$ chain with
${\theta}_{l}=0$, $q=1$ and ${s}_{l}=s>0$. In particular, for this
case, it is known that all the Bethe roots $\{x_{j}\}$ lie on the real
axis \cite{BGK1}. The number $N$ of Bethe roots corresponds to the
number of magnons in the corresponding state.
The ferromagnetic vacuum of the spin chain is defined as an 
empty state $N=0$, i.e., $Q(x)=\text{constant}$. It implies from 
the Baxter equation (\ref{baxter}) that 
\begin{align}
  - h a(x) + (h+2) d(x) = t(x) \ , 
\end{align}
which is nothing but the condition at the root of baryonic Higgs branch 
(\ref{RH}). The ferromagnetic vacuum of the
spin chain can therefore be identified as the Higgs branch root.
The eigenvalues of the transfer matrix can easily be written in terms of
the solutions of the BAE using the Baxter equation (\ref{baxter}).
\paragraph{}
In the following it will also be important that the BAE corresponds to
stationary points of an action function,
\bea
\mathbb{Y}(x) & = & 2\pi i{\tau} \, \sum_{j=1}^{N}\, x_{j}
\,\,+\,\, \sum_{j=1}^{N}\, \sum_{l=1}^{L} \, \Big[
f\left(x_{j}-\theta_{l} +is_{l}\hbar\right)
\,\,-\,\,
f\left(x_{j}-\theta_{l} -is_{l}\hbar\right) \Big] \nn \\
& &
\,\,+\,\,\sum_{j,k=1}^{N}\,
f\left(x_{j}-x_{k}+i\hbar\right)\,\,+\,\,
i\pi (N+1) \sum_{j=1}^N x_j \ , \nn \eea
where $f(x)=x(\log x\,-\,1)$ and ${q} =  \exp(2\pi i {\tau})$.
\paragraph{}
To understand the relation to the classical system of subsection 2.1,
it will be useful to consider the semiclassical limit
$\hbar\rightarrow 0$ of the
quantum chain in which the excitation numbers at each site become
large. The quantum numbers $s_{l}$ which determine the spin
representation at each site scale like $\hbar^{-1}$.
They are related to the
classical Casimirs as $J_{l}\simeq s_{l}\hbar + O(\hbar)$.
Alternatively, if $s_{l}$ are held fixed as $\hbar\rightarrow 0$
one ends up with the classical Heisenberg magnet of spin $\vec{J}=0$.
In either limit the quantum
charges $\hat{q}_{l}$ go over to Poisson commuting
classical quantities $q_{l}$.
\paragraph{}
The key feature of the semiclassical limit is that the
Bethe roots $\{x_{j}\}$ condense to form cuts in the complex plane. In
this limit the resolvent $L(x)=d\log Q(x)/dx$ is naturally defined on
a double cover of the $x$-plane with two sheets joined along these
cuts. The resulting Riemann surface is exactly the the spectral curve
$\Gamma_{L}$ and the meromorphic differential $L(x)dx$ can be
identified with the Seiberg-Witten differential $\lambda_{SW}$.
For the case of the homogeneous untwisted chain of spin zero, the
semiclassical limit is described in detail in Section 2.2 of
\cite{BGK1}. In this case the Bethe roots condense to form branch cuts
on the real axis.
Working in the vicinity of the ferromagnetic vacuum,
the curve can be represented as a double cover of the
$x$ plane with $2L$ real branch points\footnote{Strictly speaking this
  picture is correct with a real twist parameter slightly different
  from unity. In the special case $q=1$, one cut degenerates and
the genus of the curve drops to $L-2$.}
at $x=\hat x_{1}\geq
\hat x_{2}\geq\ldots\geq \hat x_{2L}$ as
shown in Figure (\ref{Qfig}). We also define one-cycles $\alpha_{l}$,
$l=1,\ldots, L$  surrounding each branch cut.
In the semiclassical limit, the quantum $SL(2,\mathbb{R})$ spin chain
gives rise to a particular real slice of the complex
classical spin chain considered above. The reality conditions select a
middle-dimensional subspace of the original complex phase space.
Allowing generic complex values of the moduli corresponds to working
with a complexification of the spin chain in which $SL(2,\mathbb{R})$
is replaced by $SL(2,\mathbb{C})$.
\paragraph{}
At the classical level, the moduli of the curve vary
continuously. The leading semiclassical approximation the quantum spectrum
arises from imposing appropriate Bohr-Sommerfeld quantisation
conditions which are formulated in terms of the periods of the
meromorphic differential $L(x)dx$ on $\Gamma_{L}$ which coincides with the
Seiberg-Witten differential $\lambda_{SW}$,
\bea
\frac{1}{2\pi} \,\oint_{{\alpha}_{l}}\,\,\lambda_{\rm SW} & =  &
\hbar\,\hat{n}_{l}  \ ,  \label{BS1} \eea
for $l=1,\ldots,L$ where $\hat{n}_{l}$ are non-negative integers. In
terms of the Bethe Ansatz, this is just the condition that each cut
contains an integral number of Bethe roots.
From this point of view it is obvious that
the quantization conditions are unchanged by continuous variations of
the parameters including the introduction of
inhomogeneities and a non-trivial twist. In
particular, they also apply in the weak coupling regime
${q}\rightarrow 0$ where the standard basis cycles of the
Seiberg-Witten curve are defined. It will be useful to express the cycles
$\vec{\alpha}$ appearing in the semiclassical quantization condition in
this basis. The key point, mentioned above, is that the ferromagnetic
vacuum of the spin chain corresponds to the Higgs branch root
$\vec{a}=\vec{m}_{F}$. In terms of the basis cycles defined above this
corresponds to the point in moduli space where the cycles
$\vec{A}-\vec{C}_{F}$ vanish. Thus we have $\vec{\alpha}=\vec{A}-\vec{C}_{F}$.

\subsection{Two-dimensional gauge theory and vortex strings}
\paragraph{}
The next observation, following \cite{NS2d}, is that the BAE
for the Heisenberg spin chain themselves arise as the F-term
equations of a certain two-dimensional gauge theory with ${\cal
N}=(2,2)$ supersymmetry which we will call Theory II. As above Theory II is
a $U(N)$ gauge theory with $L$ fundamental chiral multiplets
${Q}_{l}$ with
twisted masses $M_{l}$ and $L$ anti-fundamental chiral multiplets
$\tilde{Q}^{l}$ with twisted
masses $\tilde{M}_{l}$. The theory also contains an adjoint chiral
multiplet $Z$ with twisted mass $\epsilon$ and has a marginal
complex coupling $\hat {\tau}=ir + \theta/2\pi$ which corresponds
to a background twisted chiral superfield.
\begin{figure}
\centering
\includegraphics[width=130mm]{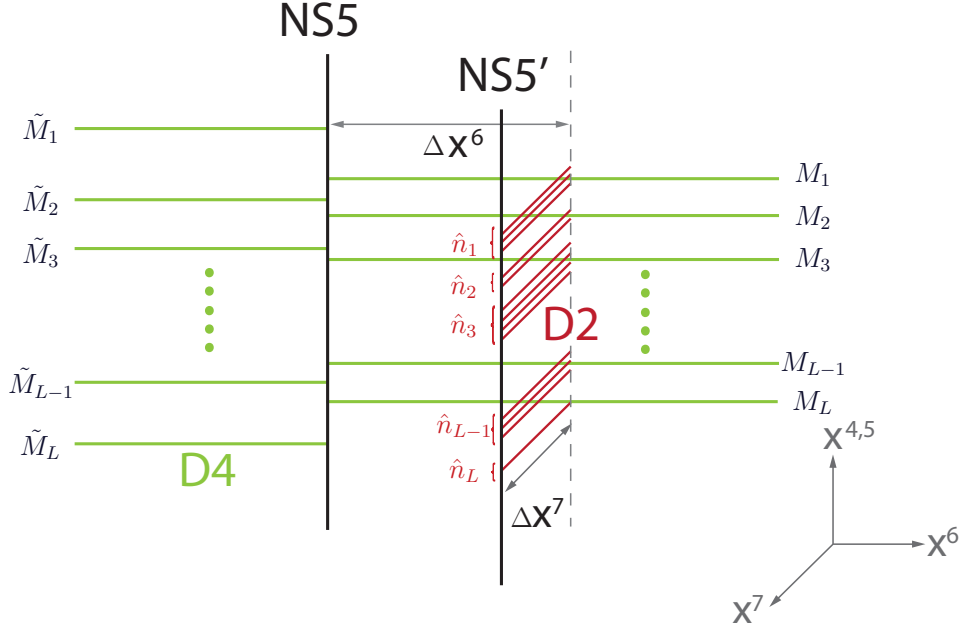}
\caption{A IIA brane construction for Theory II with $\epsilon=0$}
\label{Qfig3}
\end{figure}
\paragraph{}
We begin by focussing on the case $\epsilon=0$.  In this case the
model can be realised on the
worldvolume of $N$ D2 branes probing a configuration of intersecting
NS5 and D4 branes in Type IIA string theory \cite{Hanany:1997vm,HT1} as
shown\footnote{More precisely Theory II arises in a particular
  decoupling limit of the brane configuration.
All conventions relating to the intersecting brane
configuration shown in Figure (\ref{Qfig3}) are
the same as in \cite{DHT}.} in Figure (\ref{Qfig3}).
Since the brane-configuration is invariant under
the rotations in \{2,3\}-, \{4,5\}- and \{8,9\}-planes,
the Theory II has, at least classically,  global symmetry groups
$U(1)_{23} \times U(1)_{45} \times U(1)_{89}$ as well as
flavor symmetry groups $SU(L)\times SU(L)$.
Here the FI parameter $r$ is proportional to $\D x^6$.
Classical vacua are determined by solving the D-term equations,
\begin{align}
  \sum_{l=1}^L \Big( Q_l {Q^l}^\dagger - {\tilde Q}_l^\dagger {\tilde Q}^l \Big)
  - \big[ Z, Z^\dagger \big] = r \ ,
\end{align}
and
%
\begin{align}\label{dt}
  \sum_{l=1}^L \Big| \l Q_l - Q_l M_l  \Big|^2 + \sum_{l=1}^L \Big| -
\tilde Q^l\l
  + \tilde Q^l\tilde M_l \Big|^2 = 0 \ ,
\end{align}
where $\l$ denotes the adjoint scalar field in the vector multiplet.
\paragraph{}
For $r=0$, $Q_{l}=\tilde{Q}_{l}=0$ and Theory II has a classical
Coulomb branch parametrized by the eigenvalues
$\{\lambda_{1},\lambda_{2}, \ldots,\lambda_{N}\}$ of the adjoint
scalar field in the $U(N)$ vector multiplet. In the figure this
corresponds to the special case where each D2 is suspended between
NS5 and NS5$^{\prime}$ and can move independently in the $x_{4}$ and
$x_{5}$ directions. On the other hand, the eigenvalues of $Z$
parameterise the position of D2-branes in the \{2,3\}-plane.
\paragraph{}
For $r>0$, the theory is on a Higgs branch with
$Q\neq 0$, $\tilde{Q}=0$. The vector multiplet VEVs are fixed by
the second D-term condition (\ref{dt}). Solutions
are labelled by the number of ways of distributing
the $N$ scalars
$\{\lambda_{j}\}$ between the $L$ values $\{M_{l}\}$.
Thus we specify a
vacuum by choosing $L$ non-negative
integers $\{\hat{n}_{l}\}$ with $\sum_{l=1}^{L}\,\hat{n}_{l}=N$. In
the brane construction these correspond to the number of D2 branes
ending on each D4 brane as shown in the figure.
\paragraph{}
The brane construction also reveals an interesting physical
relation between Theory I and Theory II \cite{HT}.
In the absence of the D2 branes, the worldvolume theory on the
intersection of the remaining branes is precisely $\mathcal{N}=2$ SQCD
with gauge group $SU(L)$ and $N_{F}=2L$ hypermultiplets with masses
$m_{l}=M_{l}$ and $\tilde{m}_{l}=\tilde{M}_{l}$ and complex gauge coupling
$\tau$. In other words, it is Theory I in the undeformed case
$\epsilon=0$. To understand the relation, compare the brane
configuration shown in Figure (\ref{Qfig1}) with the configuration in
Figure (\ref{Qfig3}) (in the absence of D2 branes). The former configuration
represents a generic point on the Coulomb branch of Theory I. To pass
to the configuration shown in Figure (\ref{Qfig3}),
we reconnect the D4 branes on NS5$^{\prime}$
and then move NS5$^{\prime}$ away from the D4 branes in the $x_{7}$
direction. The first step corresponds to moving on the Coulomb branch
to the Higgs branch root root (see Figure (\ref{Qfig2})),
the second to moving out along the
Higgs branch. The theory on the Higgs branch admits vortex strings
which corresponds to the D2 branes in the Figure (\ref{Qfig3}).
Thus we identify Theory II as
the worldvolume theory on $N$ vortex strings probing the Higgs branch
of Theory I. For a review of such non-abelian vortex string see
\cite{RevVS}. 
\paragraph{}
At this point there are several subtleties. First, to have BPS
vortices with a supersymmetric worldvolume theory we must consider a
four-dimensional theory with gauge group $U(L)$ rather than
$SU(L)$. Thus, although the F-term duality described in this paper works the
same for both theories, the interpretation in terms of vortex strings
is only available in the $U(L)$ case. Next, the gauge theory
designated as Theory II above cannot be directly identified as the
worldvolume theory of the vortex. Rather, 
Theory II is an ${\mathcal N}=(2,2)$ gauged linear
$\sigma$-model which flows in the IR to a non-linear $\sigma$-model
(with twisted mass terms) whose target space is a certain K\"{a}hler manifold 
which is closely related to the moduli space of $N$ vortices 
in Theory I \cite{HT1}. In the present case, where the
four-dimensional theory is conformal, the target space has zero
first Chern class and can therefore be expected to admit a unique
Ricci-flat metric which provides a natural IR fixed-point for the worldsheet
theory. In fact, the actual K\"{a}hler metric on the classical
vortex moduli space differs from this target space metric. Even worse,
as we are discussing semi-local vortices, some elements of the true
classical moduli space metric diverge due to the non-normalisability of
some of the vortex zero modes (see eg \cite{SY3,SVY}). 
Although this has not been analysed in
detail, we expect that this problem is cured once the 
$\Omega$-background is reintroduced\footnote{An alternative way of
  regulating this divergence is proposed in \cite{SVY}}. 
Indeed, as above, a Nekrasov
deformation in one plane renders even the vacuum 
moduli of the four-dimensional gauge theory normalisable as fields in
a two-dimensional effective action. Correspondingly we expect that, in
the present context, the true world-volume theory of $N$ vortices flows 
to the same IR fixed point as Theory II. However, what we actually
need here is much weaker: that the two theories agree at the
level of $\mathcal{N}=(2,2)$ F-terms and we will assume this is the
case.     
\paragraph{}
With the identification described above, 
the numbers $\{\hat{n}_{l}\}$ labelling the classical
vacua of the worldvolume theory
determine the number of units of magnetic flux in each $U(1)$ of
the Cartan subalgebra of $U(L)$. The tension of the vortex strings is
controlled by the Higgs branch VEV $v$ of the four-dimensional
theory, proportional to $\D x^7$.
In the limit, $v \rightarrow \infty$, the tension diverges and
the vortex strings become static surface operators of the type
considered in \cite{AGGTV}.
\paragraph{}
Next we restore the Nekrasov deformation to Theory I.
From a two-dimensional perspective this corresponds to a twisted mass
term for fields charged under rotations in the $x_{2}$-$x_{3}$ plane.
The $U(N)$ adjoint chiral multiplet has charge $+1$ under this
symmetry and this field acquires mass $\epsilon$.
Thus the adjoint chiral mass in
Theory II is identified with the Nekrasov deformation parameter
$\epsilon$ in Theory I. More generally, we can consider a twisted mass
corresponding to $2$-$3$ rotations mixed with the other $U(1)$ global
symmetries which act on the chiral multiplets $Q$ and $\tilde{Q}$.
In fact the relevant symmetry for the duality we consider turns out to
be one under which $Q$ has charge $-3/2$ and $\tilde{Q}$ has charge
$+1/2$. The corresponding dictionary between the
2d and 4d masses then becomes,
\begin{align}
  \vec{M}_{F} = \vec{m}_{F} - \frac32 \vec{\e} \ ,
  \qquad
  \vec{{M}}_{{AF}} = \vec{{m}}_{{AF}} + \frac12 \vec{\e}\ .
\label{ident2} \end{align}
where $\vec{\epsilon}=(\epsilon,\e,\ldots,\e)$.
\paragraph{}
Having discussed the classical behaviour of Theory I, we now turn our
attention to the quantum theory. Quantum effects lift the
classical Coulomb and Higgs branches described above leaving only
isolated supersymmetric vacua corresponding to stationary
points of the twisted superpotential. Upon integrating out the matter
multiplets we obtain an effective twisted superpotential on the
Coulomb branch of the form,
\bea
\mathcal{W}^{(II)}(\lambda) & = &
2\pi i\hat {\tau} \, \sum_{j=1}^{N}\, \lambda_{j}
\,\,-\,\, \e \sum_{j=1}^{N}\, \sum_{l=1}^{L} \,
f\left(\frac{\lambda_{j}- M_{l}}{\e}\right) \nn \\
& & \qquad{}
\,\,+\,\, \e
\sum_{j=1}^{N}\, \sum_{l=1}^{L} \,
f\left(\frac{\lambda_{j}-\tilde{M}_{l}}{\e} \right) \,\,+\,\,\e
\sum_{j,k=1}^{N}\,
f\left(\frac{\lambda_{j}-\lambda_{k}-\epsilon}{\e} \right) \ ,
\label{spII} \eea
where $f(x)=x(\log x\,-\,1)$ and $\hat \t = i r + \frac{\th}{2\pi}$
denotes the two-dimensional holomorphic
coupling constant. It is known that this twisted
superpotential is one-loop exact. Defining
\bea
{q} =  \exp(2\pi i {\tau}) \equiv
(-1)^{N+1} \exp(2\pi i \hat \t )\ ,
\nn \eea
the resulting F-term equations can be expressed as below
\bea
\prod_{l=1}^{L}\left(\frac{\lambda_{j}-M_{l}}{\lambda_{j}-\tilde{M}_{l}}\right)
& = & {q}\, \prod_{k\neq j} \left(
\frac{\lambda_{j}-\lambda_{k}-\epsilon}{\lambda_{j}-
\lambda_{k}+\epsilon}\right)
\label{BAE2} \eea
with $j=1,2,\ldots,N$, coincide with the BAE (\ref{bae})
with the identification of the variables $\{\lambda_{j}\}$
with $\{x_{j}\}$ and setting,
\bea
\vec{M}_{F}=\vec{\theta} \,+\,i\vec{s}\hbar & \qquad{} & \vec{M}_{AF}=
\vec{\theta}\,
-\,
i\vec{s}\hbar
\nn
\eea
and $\epsilon=-i\hbar$.
The marginal coupling $\tau$ is
identified with the corresponding parameter in the Yang-Yang
functional.
\paragraph{}
The space of supersymmetric
ground states of Theory II is now identified with the Hilbert space of
the quantum $SL(2,\mathbb{R})$ spin chain.
Strictly speaking this correspondence holds when
the complex parameters of Theory II satisfy the reality conditions
implied by the above identifications. More generally one must consider an
analytic continuation of the  quantum spin chain to complex values of
its parameters. The rank $N$ of the 2d gauge group corresponds
to the number of magnons in the spin chain state. Pleasingly the total
number of states of the spin chain containing $N$ magnons is the
number of  partitions of $N$ into $L$ non-negative integers which is
the same as the number of classical SUSY vacua of the theory. It is
important to note that these vacua, like the states of the spin chain, 
correspond to {\em non-degenerate}
solutions of (\ref{BAE2}) where $\lambda_{i}\neq \lambda_{j}$ for all
$i\neq j$. Degenerate solutions are points where the Coulomb branch
effective description breaks down and any corresponding vacuum states
would have unbroken non-abelian gauge symmetry. The absence of such vacua is
consistent with the results of \cite{HoriTong}.   
\paragraph{}
To summarise the above discussion, Theory II corresponds to the
worldvolume theory of $N$ vortex strings probing the Higgs branch of
Theory I. The tension of these strings is controlled by the Higgs
branch VEV of the four-dimensional theory which corresponds to a
D-term in $\mathcal{N}=(2,2)$ superspace \cite{HT}.
The F-term equations of motion of
Theory II coincide with the BAE for the
$SL(2,\mathbb{R})$ Heisenberg spin chain. This conclusion is
independent of D-term couplings and therefore hold equally well in the
limit of infinite tension where the vortex string becomes a surface
operator.

\subsection{The duality proposal}
\paragraph{}
As mentioned above starting from a complex classical integrable system
there can be many inequivalent reality conditions and many different
quantizations. In subsections 2.1 and
2.3 we have discussed the particular reality
conditions which lead to the $SL(2,\mathbb{R})$ Heisenberg spin chain
and the standard quantization which leads to an integrable quantum spin
chain. On the other hand we have reviewed the Nekrasov-Shatashvili
quantization which produces a family of reality conditions and
quantizations related by electric-magnetic duality transformations.
It is natural to ask whether one of this family corresponds to the
standard quantization of the $SL(2,\mathbb{R})$ spin chain. To start
with we can address this question in the semiclassical regime of small
$\epsilon$. In this case the quantization conditions of the NS
proposal take the
generic form,
\bea \oint_{\vec{\mathcal{A}}}\,\,\lambda_{\rm SW} & \in & \epsilon\,\,
\mathbb{Z}\ , \nn \eea
for some choice of $L$ one-cycles $\vec{\mathcal{A}}
=(\mathcal{A}_{1},\ldots,\mathcal{A}_{L})$ on $\Gamma_{L}$
which are the image under
the low-energy duality group of the
basis cycles $\vec{B}$ appearing in the Type A quantization
condition. We obtain agreement with the semiclassical limit of the
$SL(2,\mathbb{R})$ chain if $\epsilon=-i\hbar$ and the basis cycles
are chosen as
$\vec{\mathcal{A}}=\vec{\alpha}=\vec{A}-\vec{C}_{F}$.
\paragraph{}
With the choices outlined above,
the NS quantization condition coincides with the
standard one at leading semiclassical order. In a generic quantum system it is
easy to imagine different quantizations which agree at leading order
in $\hbar$. However, for a real classical integrable system with a
given symplectic form,
quantizations which preserve integrability are very special and we do
not know of an example where two distinct quantizations coincide at
leading semiclassical order. For this reason, we conjecture that the
agreement between the NS quantization and the standard one
persists to all orders.
\paragraph{}
The Hilbert space resulting from the NS quantization corresponds to
the space of SUSY vacua determined by the superpotential,
\bea
\hat{\mathcal{W}}_{D}^{(I)}\left(\vec{a}_{D}\right) & = &
\mathcal{W}^{(I)}\left(\vec{a}\right)\,\,- \,\,\frac{2\pi i}{\epsilon}
\,\left(\vec{a}-\vec{m}_{F}\right)\cdot\vec{a}^{D}\ , \label{leg3} \eea
which is obtained by applying a duality transformation to
the superptotential $\mathcal{W}^{(I)}$ given in (\ref{wa}). In
addition to the standard electric-magnetic duality, the transformation
includes the shift $\vec{a}\rightarrow \vec{a}-\vec{m}_{F}$, 
$\vec{a}_{D}\rightarrow\vec{a}_{D}$ which is
also part of the low-energy duality group \cite{SW2}. 
The dual superpotential leads to the F-term equations,
\bea
\vec{a}-\vec{m}_{F}  & \in & \epsilon\, \mathbb{Z}^{L}\ , \nn \eea
while the Hilbert space of the standard quantization coincides with the
space of SUSY vacua with superpotential (\ref{spII}) whose F-term
equations coincide with the BAE. Equivalence between
the two quantization schemes leads to the following duality
conjecture,
\paragraph{}
{\bf Proposal} The twisted chiral rings of Theory I and Theory II are
isomorphic. This means in particular that the stationary points of the
twisted superpotentials (\ref{leg3}) and (\ref{spII}) are in one-to-one
correspondence and, with appropriate identifications between the
complex parameters, these superpotentials take the same value in
corresponding vacua;
\bea
\begin{array}{c} \\ \mathcal{W}^{(I)} \end{array} &
\begin{array}{c} _{\rm{on-shell}} \\ {\equiv}
\end{array} & \begin{array}{c} \\ \mathcal{W}^{(II)} \end{array} \nn
\eea
where equality holds up to an additive vacuum-independent
constant. Here, we the fact that $\mathcal{W}^{(I)}$ and
$\hat{\mathcal{W}}_{D}^{(I)}$ are equal on-shell as multi-valued
functions. In the next section we will test this proposal by explicit
calculation on both sides.
\paragraph{}
To complete the map between the chiral rings of the two theories we
should also give formulae for the VEVs of the
tower of chiral operators\footnote{In general, the
definition of these quantities is afflicted by the usual ambiguities
in parametrising the Coulomb branch \cite{DKM}. For fixed $L$, the
ambiguity corresponds to a finite number of {\em vacuum independent} 
coefficients.} $\hat{O}_{k}
={\rm Tr}\,\varphi^{k}=\sum_{l=1}^{L}\phi_{l}^{k}$
in each vacuum state in terms of the
corresponding set of Bethe roots $\{\lambda_{j}\}$.  
Given the correspondence between these operators and the conserved
charges of the classical spin chain which holds for $\epsilon=0$,
it is natural to conjecture that,
in the deformed theory, they are related to the
corresponding conserved charges of the quantum spin chain;
\bea
\left\langle \{\lambda_{j}\}\left| \prod_{l=1}^{L}\left(\lambda -
    \phi_{l}\right) \right| \{\lambda_{j}\}\right\rangle & = &
\frac{1}{2}t(\lambda-\lambda_{0})
\label{frst}
\eea
with,
\bea
t(\lambda) & = & \frac{-h\prod_{l=1}^{L}\left(\lambda
    -\tilde{M}_{l}\right)\prod_{j=1}^{N}
\left(\lambda-\lambda_{j}-\epsilon\right)\,\,+\,\,(h+2)\prod_{l=1}^{L}\left(
\lambda-{M}_{l}\right)\prod_{j=1}^{N}
\left(\lambda-\lambda_{j}+\epsilon\right)}{ \prod_{j=1}^{N}
\left(\lambda-\lambda_{j}\right)}
\nn
\eea 
Here the shift parameter $\lambda_{0}$ is chosen to impose the $SU(L)$
condition $\sum_{l=1}^{L}\phi_{l}=0$. In the weak-coupling limit
$\tau\rightarrow\infty$, one may check the formula holds
with $\lambda_{0}=\epsilon/2$ using the results of the next section.
These additional predictions will be discussed further in \cite{us2}.
\paragraph{}
Finally we note that other
aspects of the relation between ${\cal N}=2$ SQCD and the
quantum spin chain
have been checked very recently in \cite{Zenk}. Specifically, in
this paper, it is verified that the A- and B-type quantization
conditions match the quantisation conditions for
periodicity of a
wave-function obeying the Baxter equation of the spin chain for
general complex values of the parameters.
A similar agreement to the quantization condition of the Gaudin model
was obtained earlier in \cite{Taki1}.

\section{A Weak-Coupling Test}

We present in this section a detailed comparison at the weak coupling limit
between the twisted superpotentials of both theories in the proposed duality.
As discussed below, a careful analysis
shows perfect agreement.
\paragraph{}
In the following we will find a correspondence where the numbers
$\hat{n}_{l}$ of D2 branes in Theory II are identified with the
integers appearing in the quantization condition
$a_{l}=m_{l}-n_{l}\e$ for Theory I according to
$\hat{n}_{l}=n_{l}-1$. Suggestively both sets of integers correspond
to quantized magnetic fluxes in the Cartan subalgebra of the gauge
group. As $\hat{n}_{l}$ are by definition non-negative $n_{l}$ must be strictly
positive. The trace condition for the $SU(L)$ gauge group also implies the
additional constraint on the sum of the hypermultiplet masses;
$\sum_{L=1}^{L}m_{l}=(N+L)\e$.

\subsection{Theory I}

The Nekrasov partition function is composed of
three factors corresponding to the classical, the
one-loop perturbative, and the instanton contributions:
\begin{align}
  \CZ_\text{Nek}(\vec a,\t,\e_1,\e_2) =
  \CZ_\text{cl} \times \CZ_\text{1-loop} \times \CZ_\text{inst} \ ,
  \nonumber
\end{align}
which leads to a corresponding decomposition
of the twisted superpotential of Theory I
\begin{align}
  \CW^{(I)}(\vec a,\t , \e) = \CW_\text{cl} + \CW_\text{1-loop} +
\CW_\text{inst}\ .
\end{align}
We will evaluate the on-shell value of these terms in order.
More precisely, we will compute a difference of the
twisted superpotential value at a vacuum $\vec a = \vec m - \vec n \e$
and at the root of the baryonic Higgs branch\footnote{
Note the shift of $-\e$ in the
location of the Higgs branch root which follows from the conventions
for the $\Omega$ background used in \cite{NO,AGT}.
The location of the Higgs branch root is
determined as the point on the Coulomb branch where additional
hypermultiplets become massless.}
$\vec a= \vec m - \e$  which corresponds to the ferromagnetic ground
state of the spin chain. This subtraction corresponds to a vacuum
independent ({\it i.e.\/}~$\vec{n}$-independent)
constant shift of the superpotential.
\begin{align}
  \CG = \left. \CW^{(I)}  \right|_{\vec a = \vec m - \vec n \e}
  - \left. \CW^{(I)} \right|_{\vec a = \vec m - \e}\ .
\end{align}
Strictly speaking, the standard results of \cite{N1,NO} apply to the
partition function of the
$U(L)$ theory which differs from the Nekrasov partition function of
the $SU(L)$ theory by a
multiplicative factor which does not depend on the Coulomb branch
moduli $\vec{a}$ \cite{AGT}. The resulting
difference between the superpotentials of the $SU(L)$ and
$U(L)$ theories is therefore vacuum-independent
additive constant which is not visible in our analysis.
\paragraph{}
In order to avoid degenerate cases,
we suppose that $n_l$, $l=1,2,\ldots,L$, satisfy the following relations
\begin{align}
  n_1 \geq n_2 \geq ... \geq n_{L} \ , \qquad
  m_l - n_l \e \geq m_n - n_n \e \qquad \text{ if } l < n\ .
\end{align}

\paragraph{Classical part:}
The classical part $\CZ_\text{cl}$ of the partition function is
given by
\begin{align}
  \CZ_\text{cl} = \exp\Big[ - \frac{2\pi i}{\e_1\e_2} \sum_{l=1}^L \frac{\t_0}{2}
   a_l^2 \Big]\ , \nonumber
\end{align}
where as above $\t_0$ denotes the bare marginal coupling constant of the
$SU(L)$ gauge group. This coupling is slightly different to the
the coupling $\t$ in the Seiberg-Witten curve (\ref{sw1}), due to
some ambiguities in the perturbative computation. We
will discuss it further below. The classical part of the function $\CG$
is therefore given by
\begin{align}\label{gcl}
  \CG_\text{cl} = &
  -\frac{2\pi i \t_0}{\e} \sum_{l=1}^L \frac12 \Big( (m_l - n_l \e)^2 -
  (m_l - \e )^2 \Big)
  \nonumber \\ = &
  2 \pi i \t_0 \sum_{l=1}^L  \Big( (n_l - 1) (m_l-\e) - \frac12 (n_l-1)^2
  \e  \Big)\ .
\end{align}

\paragraph{1-loop part:}
The one-loop contribution $\CZ_\text{1-loop}$ consists of three factors
coming from the vector multiplet and $L$ fundamental/anti-fundamental
hypermultiplets. As mentioned in \cite{N1} and above,
the one-loop contribution has an ambiguity in fixing the quadratic terms.
It describes the finite renormalization of the classical part of $Z$.
Following the convention of \cite{NO},
\begin{align}
  \CZ_\text{1-loop} = z_\text{vec} \times z_\text{fund} \times
  z_\text{anti-fund}\ ,  \nonumber
\end{align}
with
\begin{align}
  z_\text{vec} = & \prod_{l,m=1}^L
  \frac{1}{\text{Exp}\big[
  \g_{\e_1,\e_2}(a_{lm} )  \big]}\ ,
  \nonumber \\
  z_\text{fund} = &
  \prod_{l,m=1}^L \text{Exp}\big[
  \g_{\e_1,\e_2}(a_l - m_m ) \big]\ ,
  \nonumber \\
  z_\text{anti-fund} = &
  \prod_{l,m=1}^L \text{Exp}\big[
  \g_{\e_1,\e_2}(a_l - \tilde m_m - \e_1 - \e_2) \big]\ .
\end{align}
where $a_{lm} = a_l - a_m$. Here, the function $\g_{\e_1,\e_2}(x)$
is the logarithm of Barnes' double gamma function
whose properties are summarized in \cite{NO}. In particular, the function
$\g_{\e_1,\e_2}(x)$ obeys a difference equation below
\begin{align}\label{relation1}
   \g_{\e_1,\e_2}(x)+ \g_{\e_1,\e_2}(x-\e_1-\e_2) - \g_{\e_1,\e_2}(x-\e_1)
   -\g_{\e_1,\e_2}(x-\e_2) = \log {\frac \L x} \ .
\end{align}
Here, we are primarily interested in its limiting form as $\e_2\to 0$.
Defining a new function $\w_\e(x)$ with $\L=\e$ by
\begin{align}
  \w_\e(x) = \lim_{\e_2 \to 0} \Big[ \e_2 \g_{\e_1=\e,\e_2} (x) \Big] \ ,
\end{align}
one can then verify from (\ref{relation1}) the following relation
\begin{align}
  \w'_\e(x) - \w'_\e(x-\e)=  \log  {\frac x\e}\ , \qquad
  \w'_\e(x) = \frac{d}{dx} \w_\e(x)\ . \nonumber
\end{align}
As in \cite{NS4d}, it immediately implies  that
\begin{align}\label{relation2}
  \w'_\e(x) = - \log \G( 1 + x/\e )\ .
\end{align}
For later convenience, we present a relation for $\w_\e(x)$ with positive $n$
\begin{align}\label{formula3}
  \w_\e ( x + n\e ) - \w_\e(x ) = - \e \sum_{j=1}^n f\big( \frac{x}{\e} + j \big)
  - \frac{n\e}{2} \log 2\pi\ , \qquad
  f(x) = x \big( \log x -1 \big)\ ,
\end{align}
which will be frequently used in what follows. The branch of the
multi-valued function $f(x)$ is chosen here that, for non-negative
$x$,
\begin{align}
  f(x) = - f(-x) + i \pi x \ . \nonumber
\end{align}

\paragraph{}
The one-loop contribution to the twisted superpotential of
Nekrasov and Shatashivili can therefore be expressed as
\begin{align}
  \CW^{(I)}_\text{1-loop} = \varpi_\text{vec} + \varpi_\text{fund}
  + \varpi_\text{anti-fund}
  \nonumber
\end{align}
with
\begin{align}
  \varpi_\text{vec} = & - \sum_{l,m} \w_\e(a_{lm}  ) \ ,
  \nonumber \\
  \varpi_\text{fund} = & \sum_{l,m}  \w_\e (a_l - m_m)\ ,
  \nonumber \\
  \varpi_\text{anti-fund} = & \sum_{l,m} \w_\e (a_l -\tilde m_m
  - \e )\ .
\end{align}
Using the relation (\ref{formula3}), one can show that
\begin{align}
  \CG_\text{1-loop}^\text{fund} = & \e \sum_{l,m} \bigg[\sum_{k=1-n_l}^{-1}
  f\big(  \frac{m_{lm}}{\e} +k\big) + \frac{n_l-1}{2} \log 2\pi \bigg]
  \nonumber \\
  \CG_\text{1-loop}^\text{anti-fund} = &  \e \sum_{l,m} \bigg[
  \sum_{k=0}^{n_l-2} f ( \frac{m_l - \tilde m_m-2\e}{\e} - k ) + \frac{n_l-1}{2}
  \log 2\pi \bigg]\ ,
  \nonumber \\
  \CG_\text{1-loop}^\text{vec} = &- \e  \sum_l \sum_{m>l} \bigg[
  \sum_{k=1+n_m-n_l}^{0} + \sum_{k=n_m - n_l }^{-1}\bigg]
  f \big( \frac{m_{lm}}{\e}+ k\big) + i\pi \e \sum_{l,m>l} \sum_{k=n_m-n_l}^{-1}
  \big( \frac{m_{lm}}{\e} + k \big) \  .
\end{align}
After some algebra whose details are discussed in Appendix A,
one can finally obtain
\begin{align}\label{gp}
  \CG_\text{1-loop} = &- \e \sum_{l,m=1}^L \sum_{k=1+n_m-n_l}^{n_m-1} f\big( \frac{m_{lm}}{\e} + k \big)
  + i \pi  L \sum_l \bigg[ (n_l -1) m_l -
\frac{(n_l-1)^2}{2}\e \bigg]\ ,
\end{align}
up to some constants which do not depend on the choice of vacuum.
\paragraph{}
Collecting the results so far, the perturbative contribution $\CG_\text{pert}
= \CG_\text{cl} + \CG_\text{1-loop}$ takes the following form
\begin{align}\label{pert}
  \CG_\text{pert}
  = 2 \pi i \t \sum_{l=1}^L  \Big( (n_l - 1) (m_l-\e) - \frac12 (n_l-1)^2
  \e  \Big)
  - \e \sum_{l,m=1}^L \sum_{k=1+n_m-n_l}^{n_m-1} f\big( \frac{m_{lm}}{\e} + k \big)\ ,
\end{align}
where $\t =\t_0 + \frac L2 $ denotes the marginal coupling constant in the curve.

\paragraph{Instanton contribution:}
The instanton contribution to the partition function is written as a
sum over coloured partitions
\bea
\mathcal{Z}_{\rm inst} & = & 1\, +\,
\sum_{k=1}^{\infty}\,\mathcal{Z}_\text{inst}^{(k)}\, \big( (-1)^L q\big)^{k} \nn \\ & =&
\sum_{\vec{Y}}\, \CZ_{\vec{Y}}\, (-1)^{L|\vec Y|}q^{|\vec{Y}|}\ , \nn
\eea
where the factors of $(-1)^L$ are included so that definition of $q=e^{2\pi i \t}$
agrees with the coupling that appears in the Seiberg-Witten curve.
For $U(L)$ gauge group, the coloured partition $\vec Y=(Y_1, Y_2,..,Y_L)$
are labelled by $N$  Young diagrams $Y_i$.
The total number of boxes $|\vec Y| = \sum_{i=1}^N |Y_i|$
is the instanton number $k$.  The explicit form of
$\CZ_{\vec{Y}}$ can be found in \cite{AGT}.

\paragraph{}
Denoting the unique partition of unity as $\{\bf 1\}$ and the
two inequivalent partitions of two as $\{\bf 2\}$ and $\{\bf 1,1\}$,
the following coloured partitions can contribute:
\begin{enumerate}
  \item[$\bullet$] In the one-instanton sector $k=1$,
  we have partitions $\vec{Y}^{(l)}_{\{\bf 1\}}$ with components
  $(Y^{(l)}_{\{\bf 1\}})_{m}=\delta_{lm}\{\bf 1\}$ for $l,m=1,2,\ldots,
  L$. The one-instanton contribution to the partition function
  therefore becomes
  \begin{align}
    \CZ^{(1)}_\text{inst} = \CZ_{\{\bf 1\}}
    = -\,\frac{(-1)^L}{\epsilon_{1}\epsilon_{2}}\,
    \sum_{l=1}^{L}\, R_{l}\left(a_l,\epsilon_1+\e_2\right)\ ,
  \end{align}
  where
  \begin{align}
    R_l\big(x,\e) =  \frac{\prod_{m=1}^L (x -m_m + \e) (x - \tilde m_m)}
    {\prod_{m\neq l} (x - a_m  + \e) (x - a_m)} \ .
  \end{align}

  \item[$\bullet$] In the two-instanton sector $k=2$, we have
  coloured partitions $\vec{Y}_{\{\bf 2\}}^{(l)}$,
  $\vec{Y}_{\{\bf 1,1\}}^{(l)}$ ($l=1,2\ldots,L$), and
  $\vec{Y}_{\{\bf 1\},\{\bf 1\}}^{(l,m)}$ ( $l,m=1,2,\ldots,L$ and $l\neq m$)
  with components
  \begin{align}
    \delta_{ln}{\{\bf2\}}\ , \ \delta_{ln} {\{\bf 1,1\}}\ ,
    \   \text{ and } \
    \delta_{l n}{\{\bf 1\}} + \delta_{m n}{\{\bf 1\}}\ , \nn
  \end{align}
  respectively. The corresponding contributions to the partition function are
  given by
  \begin{align}
    \CZ_{\{\bf 2\}} = & \
    \frac{1}{2\epsilon_{1}\epsilon_{2}^{2}
    (\epsilon_{1}-\epsilon_{2})}\, \sum_{l=1}^{L} \
    R_{l}\left(a_l, \epsilon_1+\e_2 \right)\
    R_{l}\left(a_l+\epsilon_{2}, \epsilon_1+\e_2 \right)\ ,
    \nonumber \\
    \CZ_{\{\bf 1,1\}} = & \
    \frac{1}{2\epsilon_{2}\epsilon_{1}^{2}
    (\epsilon_{2}-\epsilon_{1})}\, \sum_{l=1}^{L} \
    R_{l}\left(a_l, \epsilon_1+\e_2 \right)\
    R_{l}\left(a_l+\epsilon_{1}, \epsilon_1+\e_2 \right)\ ,
    \nonumber \\
    \CZ_{\{\bf 1\} , \{\bf 1\} } = & \
    \frac{1}{\epsilon_{1}^{2}\epsilon_{2}^{2}}
    \sum_{l\neq m} R_{l}\left(a_l,\epsilon_1+ \e_2 \right)\,
    R_{m}\left(a_m,\epsilon_1 + \e_2 \right) \,\,
    \frac{a_{lm}^{2}\left(a_{lm}^{2} -(\epsilon_1+\e_2) ^{2}\right)}
    {\left(a_{lm}^{2} -\epsilon_{1}^{2}\right)\left(a_{lm}^{2}
    -\epsilon_{2}^{2}\right)}\ .
  \end{align}
 The two-instanton contribution to the partition function
takes the following form
  \begin{align}
    \CZ_\text{inst}^{(2)} = \CZ_{\{\bf 2\}} + \CZ_{\{\bf 1,1\}}
    + \CZ_{\{\bf 1\}, \{\bf 1\}}\ .
  \end{align}
\end{enumerate}
Since the logarithm of the instanton partition function
can be expanded as below
\begin{align}
  \log  \CZ_\text{inst} = &
 (-1)^L q \CZ_\text{inst}^{(1)} + q^2 \bigg[ \CZ_\text{inst}^{(2)}
  - \frac12 \big( \CZ_\text{inst}^{(1)} \big)^2 \bigg] + \CO (q^3)
  \nonumber \\ = &
  (-1)^Lq \CZ_{\{\bf 1\}} + q^2 \bigg[ \CZ_{\{\bf 2\}} + \CZ_{\{\bf 1,1\}}
  + \CZ_{\{\bf 1\}, \{\bf 1\}} - \frac12 \big( \CZ_{\{\bf 1\}} \big)^2
  \bigg] + \CO(q^3)\ ,  \nonumber
\end{align}
the instanton contribution to the twisted superpotential
becomes
\begin{align}
  \CW^{(I)}_\text{inst} = & \lim_{\e_2\to 0} \big[ \e_2
  \log  \CZ_\text{inst} ( \e_1=\e, \e_2 ) \big]
  =  \sum_{k=1}^\infty \CW^{(k)}_\text{inst} q^k\ ,
\end{align}
where
\begin{align}
  \CW^{(1)}_\text{inst} = & - \frac1\e \sum_{l=1}^L R_l ( a_l,\e)
  \nonumber \\
  \CW^{(2)}_\text{inst} = & + \frac{1}{2\e^3} \sum_{l=1}^L
  \bigg[ R_l(a_l,\e)^2 - R_l(a_l,\e) R_l(a_l+\e,\e) + \epsilon R_l(a_l,\e)
  {R}'_l(a_l,\e) \bigg]
  \nonumber \\
  & - \frac1\e \sum_{l\neq m} \frac{1}{a^2_{lm} - \e^2}
  R_l(a_l,\e) R_m(a_m,\e)\ .
\label{gts}
\end{align}
%
%
\paragraph{}
Due to the fact that there arise additional zero modes at the
root of baryonic branch ,
the instanton contribution to the partition function
should vanish at $\vec a = \vec m -\e$. Indeed one can show that
$$
R_l(m_l-\e) = 0 \ .
$$
The above relation simplifies the evaluation of
the instanton contribution to the function $\CG_\text{inst}$ of
our interest, and finally we evaluate the superpotential on-shell to
obtain,
\begin{align}
  \CG_\text{inst} = \sum_{k=1}^\infty \CG_\text{inst}^{(k)} q^k
  \nonumber
\end{align}
with
\begin{align}\label{ginst}
  \CG^{(1)}_\text{inst} = &
  \left. \CW^{(1)}_\text{inst} \right|_{a_l = m_l -n_l \e}
  =   - \frac1\e \sum_{l=1}^L R_l ( m_l - n_l\e)  \ , \nonumber \\
  \CG^{(2)}_\text{inst} =  & \left. \CW^{(2)}_\text{inst}
  \right|_{a_l=m_l - n_l \e}
   \\ = &
  + \frac{1}{2\e^3} \sum_{l=1}^L
  \bigg[ R_l(m_l-n_l\e,\e)^2 - R_l(m_l-n_l\e ,\e)
  R_l(m_l - (n_l-1) \e,\e)
  \nonumber \\
  & \hspace*{2cm} + R_l(m_l-n_l\e,\e)  {R'}_l(m_l-n_l\e,\e) \bigg]
  \nonumber \\
  & - \frac1\e \sum_{l\neq m} \frac{1}{\big(m_{lm}-(n_l -n_m)\e\big)^2 - \e^2}
  R_l(m_l-n_l\e,\e) R_m(m_m-n_m\e,\e)\ .
  \nonumber
\end{align}

\paragraph{}
\subsection{Theory II}
Let us now consider the Theory II as introduced in subsection 2.4.
The one-loop exact twisted superpotential $\CW^{(II)}$
on the Coulomb branch takes the following form
\begin{align}
  \CW^{(II)} = \CW_\text{tree} + \CW_\text{fund}
  + \CW_\text{anti-fund} + \CW_\text{adj} \ ,
\end{align}
where
\begin{align}\label{2D}
  \CW_\text{tree} = &  2\pi i \t \sum_{i=1}^{N}
  \lambda_i - i \pi (  N + 1 ) \sum_{i=1}^{N} \lambda_i
  \nonumber \\
  \CW_\text{fund} = & - \e \sum_{i=1}^{N} \sum_{l=1}^{L}
  f\big( \frac{\lambda_i -  M_l}{\e} \big)
  \nonumber \\
  \CW_\text{anti-fund} = &  \e
  \sum_{i=1}^{N} \sum_{l=1}^{L} f\big( \frac{\lambda_i - \tilde M_l}{\e}\big)
  \nonumber \\
  \CW_\text{adj} = & \e \sum_{i,j=1}^{N} f\big(
  \frac{\lambda_i - \lambda_j}{\e} -1 \big)\
\end{align}
with $f(x)=x(\log x - 1)$.
Here $\t$ is related to the two-dimensional holomorphic
coupling constants $\hat \t = ir + \th/2\pi$ as
$\hat \t = \t - \frac{N+1}{2}$.

\paragraph{Vacuum solution:} As discussed before, it turns out that the
F-term vacuum equation $\exp(
\partial \CW^{(II)}/\partial \lambda_i)= 1$
indeed leads us to an algebraic BAEs
\begin{align}\label{bethe}
  \prod_{l=1}^L  \big( \lambda_i - M_l \big) \prod_{j\neq i}
  \big( \lambda_i - \lambda_j + \e \big) =  q
  \prod_{l=1}^L \big( \lambda_i - \tilde M_l \big)
  \prod_{j\neq i} \big( \lambda_i - \lambda_j - \e \big)\ ,
%
\end{align}
where $q= e^{2\pi i \t}$. One can solve the above
BAE in an iterative way by expanding the lowest component of
twisted chiral superfield $\lambda_i$ in the instanton factor $q$
\begin{align}\label{exp}
  \lambda_i = \sum_{k=1}^\infty \lambda_i^{(k)} q^k \ .
\end{align}
\paragraph{}
At leading order, we have
\begin{align}
  \prod_{l=1}^L \big( \lambda_i^{(0)} - M_l \big) \prod_{j\neq i}
  \big( \lambda_i^{(0)} - \lambda_j^{(0)} + \e \big) = 0\ .
\end{align}
Obeying the non-degeneracy condition $\lambda_i^{(0)} \neq \lambda_j^{(0)}$,
one can show that the solution at leading order can be characterized
by partition of $N$ with $L$ parts, i.e.,
\begin{align}
  \big\{ \hat n_1,\hat n_2,..\hat n_L \big\}\ ,
  \qquad \sum_{l=1}^L \hat n_l = N\ ,
\end{align}
and takes the following form
\begin{align}\label{zero}
  \lambda_i^{(0)} = \lambda^{(0)}_{(l,s_l)} = M_l - (s_l -1 ) \e\ ,
\end{align}
where $s_l$ runs over $1,2,..,\hat n_l$.
In the language of the spin chain these solutions take the form of
Bethe strings. Here the strings arise in an unfamiliar limit where the
twist parameter $h\simeq -2 q$ goes to zero. Such strings are
more usually associated with the
thermodynamic limit of the spin chain where they
represent magnon bound states.

\paragraph{}
Let us now in turn discuss the solution at the next order $\lambda^{(1)}_i$.
Expanding the equation (\ref{bethe}) to the next order, one
can show that $\lambda^{(1)}_{(l,s_l)}$ always vanish except $s_l = \hat n_l$.
One can thus set
\begin{align}
  \lambda_{(l,s_l)}^{(1)} = \d_{s_l, \hat n_l} \lambda_l^{(1)}\ ,
\end{align}
where $\lambda_l^{(1)}$ satisfies the relation below
\begin{align}
  \bigg [ \prod_{m=1}^L \big( M_l -M_m - (\hat n_l -1 ) \e \big)
  \prod_{(m,s_m) \neq (l, \hat n_l),(l, \hat n_l -1)} \big( M_l - M_m
  - (\hat n_l - s_m -1 )\e \big) \bigg] \cdot \lambda_l^{(1)}
  \nonumber \\
  =
  \prod_{m=1}^{L} \big(M_l -  \tilde M_m - (\hat n_l -1) \e  \big)
  \prod_{(m,s_m) \neq (l,\hat n_l)} \big( M_l - M_m - ( \hat n_l -s_m +1 )
  \e \big)\ .
\end{align}
After some algebra, it leads to
\begin{align}\label{first}
  \lambda_l^{(1)} = & - \frac{1}{\e} \cdot
   \frac{\prod_{m} \big(M_{lm} -  \hat n_l \e \big) \cdot
   \big(M_l -\tilde M_m - (\hat n_l - 1) \e \big)}
   {\prod_{m\neq l} \big(M_{lm} - ( \hat n_l -\hat n_m ) \e \big)
   \cdot \big(M_{lm} - (\hat n_l - \hat n_m -1) \e \big) }\ .
\end{align}
\paragraph{}
Expanding (\ref{bethe}) to the next order we discover that
$\lambda^{(2)}_{l,s_l}$ again vanishes unless $s_{l}=\hat{n}_{l}$ or
$\hat{n}_{l}-1$.
The expansion (\ref{exp}) therefore takes the following form,
\begin{align}
  \lambda_{l,s_l} = \lambda^{(0)}_{l,s_l }+  \d_{s_l, \hat n_l} \lambda_l^{(1)}  q
  + \Big( \d_{s_l,\hat n_l} \lambda_l^{(2)} + \d_{s_l, \hat n_l-1}
  \tilde \lambda_l^{(2)} \Big)q^2 + \cdots\ .
  \nonumber
\end{align}
As explained later, we will only need an explicit expression for
$\tilde{\lambda}^{(2)}_{l}$ and not $\lambda^{(2)}_{l}$.
Setting $s_{l}=\hat{n}_{l}-1$ in (\ref{bethe}) and keeping
all terms of order $q^{2}$, one can read the
resulting equation that $\tilde \lambda_l^{(2)}$ should satisfy
\begin{align}
  \bigg[ \prod_{m=1}^L & \big( M_{lm} - (\hat n_l -2 )\e \big)
  \prod_{(m,s_m)\neq( l ,\hat n_l-1) , (l ,\hat n_l -2)} \big( M_{lm}
  - ( \hat n_l -s_m -2 )\e \big)  \bigg] \cdot  q^2 \tilde \lambda_{l}^{(2)}
  \nonumber \\ = &
  q \bigg[ \prod_{m=1}^{L} (M_l -  \tilde M_m - (\hat n_l -2) \e  )
  \prod_{(m,s_m) \neq (l,\hat n_l-1), (l ,\hat n_l)}
  ( M_{lm} - ( \hat n_l -s_m)\e  )\bigg] \cdot (- q \lambda^{(1)}_l )  \ ,
\end{align}
which yields
\begin{align}\label{second}
  \tilde \lambda_l^{(2)} = &  \frac{\lambda^{(1)}_l }{2\e^2} \cdot
  \frac{\prod_{m=1}^L \big( M_{lm} - (\hat n_l -1)\e \big)
  \big(M_l -\tilde M_m - (\hat n_l -2 )\e  \big) }
  {\prod_{m\neq l} \big(M_{lm} - ( \hat n_l -\hat n_m -1)\e \big)
  \big( M_{lm} - (\hat n_l - \hat n_m -2) \e \big) }\ .
\end{align}
\\
We are now ready to evaluate the on-shell value of
the twisted superpotential $\CW^{(II)}$ up to
the lowest three orders in the instanton factor $q$
\begin{align}
  \CW^{(II)} =
  \CW^{(0)} + \CW^{(1)} q + \CW^{(2)} q^2 + \CO( q^3)\ .
\end{align}
\paragraph{Leading order, $\CW^{(0)}$:} Substituting the solution
at leading order $\lambda_{m}^{(0)}$ (\ref{zero}) to (\ref{2D}), one can
have
\begin{align}
  \CW^{(0)} = \CW_\text{tree}^{(0)} + \CW_\text{fund}^{(0)}
  + \CW_\text{anti-fund}^{(0)} + \CW_\text{adj}^{(0)}\ ,
  \nonumber
\end{align}
with
\begin{align}
  \CW_\text{tree}^{ (0)} = & \big(2\pi i  \t -  i \pi (N+1) \big) \sum_{l=1}^{L}
  \sum_{s_l=1}^{\hat n_l} \big( M_l - (s_l-1) \e \big)
  \nonumber \\ = &  \big(2\pi i  \t -  i \pi (N+1) \big)\sum_{l=1}^L
  \big( \hat n_l (M_l + \frac{\e}{2} ) - \frac{\hat n_l^2}{2} \e \big)\ ,
\end{align}
and
\begin{align}
  \CW_\text{fund}^{(0)} = & - \e \sum_{l,m=1}^L \sum_{k=0}^{\hat n_l-1}
  f\big( \frac{M_{lm}}{\e} - k\big)\ ,
  \nonumber \\
  \CW_\text{anti-fund}^{(0)} = & + \e \sum_{l,m=1}^L \sum_{k=0}^{\hat n_l-1}
  f\big( \frac{M_l - \tilde M_m}{\e} - k \big)\ , \nonumber \\
  \CW_\text{adj}^{(0)} = & + \e \sum_{l,m=1}^L \sum_{s_l=1}^{\hat n_l}
  \sum_{s_m=1}^{\hat n_m} f \big( \frac{M_{lm}}{\e}
  - s_l + s_m -1 \big) \ .
\end{align}
One can show that, after some algebra presented in Appendix A,
the twisted superpotential at leading order $\CW^{(0)}$ can be
simplified into the following form
\begin{align}\label{wzero}
  \CW^{(0)} = &
  2\pi i \t \sum_{l=1}^L
  \big( \hat n_l (M_l + \frac{\e}{2} ) - \frac{\hat n_l^2}{2} \e \big)
  - \e \sum_{l, m=1}^L \sum_{k=\hat n_m -\hat n_l +1}^{\hat n_m}
  f\big( \frac{M_{lm}}{\e} + k \big) \ ,
\end{align}
again up to some constant independent of the vacuum choice.
We also used properties of the multi-valued function $f(x)$ extensively.

\paragraph{Next two leading orders, $\CW^{(1)}$ and $\CW^{(2)}$:}
Using the vacuum equation, it is straightforward to obtain
the next two leading contribution to the twisted superpotential
as below
\begin{align}
  \CW^{(1)} = \sum_{l=1}^L \lambda_l^{(1)}
\end{align}
and
\begin{align}
  \CW^{(2)} & =  \sum_{l=1}^L
  \tilde \lambda_l^{(2)} +  \frac{1}{2\e} \big( \lambda_l^{(1)} \big)^2 \,+\,
  \sum_{l=1}^L \big( \lambda_l^{(1)} \big)^2 \bigg[
  \sum_{m=1}^L \Big( \frac{1}{M_{lm}-\hat{n}_{m}\epsilon} +
  \frac{1}{M_{l}-\tilde{M}_{m}-\hat{n}_{l}\epsilon+\epsilon} \Big)
  \nonumber \\ & \hspace*{0.5cm}
  - \sum_{m\neq l} \Big(\frac{1}{M_{lm}-(\hat{n}_{l}-\hat{n}_{m})\epsilon}
  + \frac{1}{M_{lm}-(\hat{n}_{l}-\hat{n}_{m} -1 )\e } \Big)
  \bigg]
  \\ & \hspace*{0.5cm}
  - \e \sum_{m\neq l}
  \frac{\lambda_l^{(1)} \lambda_m^{(1)} }{\big( M_{lm} - (\hat n_l -\hat n_m)\e\big)^2 - \e^2}
  \nonumber\ .
\end{align}
Here $\lambda_l^{(1)}$ and $\lambda_l^{(2)}$ are explicitly given by
(\ref{first}) and (\ref{second}), respectively.

\subsection{Comparison}

In order to compare the two theories, we first identify parameters
in both theories as follows
\begin{align}
  \hat n_l = n_l - 1 \ ,
  \qquad
  M_l = m_l - \frac32 \e \ ,
  \qquad
  \tilde M_l = \tilde m_l + \frac12 \e\ ,
\end{align}
Upon the above identification, one can compare
the on-shell twisted superpotentials of both theories
order by order in instanton factor $q$.
One can find the agreement at leading order
between (\ref{pert}) and (\ref{wzero}), i.e.,
\begin{align}
  \CG_\text{pert} = \CW^{(0)}\ .
\end{align}
Noting that
\begin{align}
  \lambda_l^{(1)} = & -\frac{1}{\e} R_l \big( m_l - n_l \e \big)\ ,
  \nonumber \\
  \tilde \lambda_l^{(2)} = & - \frac{1}{2\e^3} R_i\big( m_l - n_l \e)
  R_l \big( m_l - (n_l -1)\e \big)\ ,
\end{align}
one can show the agreement further  up to the two-instanton sector
\begin{align}
  \CG_\text{inst}^{(1)} = \CW^{(1)} \ , \qquad
  \CG_\text{inst}^{(2)} = \CW^{(2)}\ .
\end{align}
%

%



\section{Discussion and Relation to Other Developments}
\begin{figure}
\centering
\includegraphics[width=140mm]{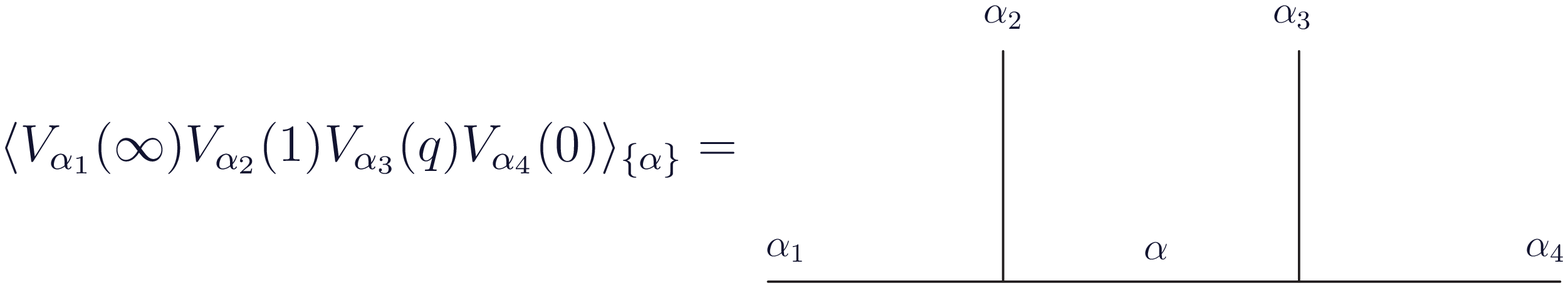}
\caption{Four-point Liouville conformal block on the sphere.}
\label{Qfig4}
\end{figure}

\paragraph{}
The duality proposed above also has several points of contact with other
recent developments in the study of supersymmetric gauge theory. In
particular we can interpret our results in the context of the AGT
conjecture which relates the Nekrasov instanton partition function of
${\cal N}=2$ supersymmetric $G=SU(2)$ gauge theory
in four dimensions to the conformal blocks of Liouville theory.  For
$L=2$, Theory I defined above is an $SU(2)$ gauge theory with
$N_{F}=4$ fundamental hypermultiplets. The electric
Coulomb branch parameters take the form $\vec{a}=(a,-a)$. According to the AGT
conjecture the Nekrasov instanton partition function
$\mathcal{Z}(\vec{a}, \epsilon_{1},\epsilon_{2})$ is related to the
conformal block for the correlation function of four primary operators
$V_{\alpha_i}(z)=\exp[2\alpha_{i}\phi(z)]$,
inserted at points $z=z_{i}$ ($i=1,2,3,4$) on the sphere.
The marginal coupling constant of the four dimensional gauge theory
$q=e^{2\pi i\t}$
can be identified with conformally invariant cross-ratio of these four points,
the single modulus of a sphere with four punctures.
Fixing three positions of vertex operators by $0,1$ and $\infty$ as usual,
the left-over one therefore parametrizes the coupling $q$.
The Liouville coupling and effective Planck
constant\footnote{This should not be confused with the Planck constant
  of the integrable system discussed in Section 2.} are given as,
\bea
b= \sqrt{\frac{\epsilon_{1}}{\epsilon_{2}}}\ , & \qquad{} \qquad{} & \hbar =
\sqrt{\epsilon_{1}\epsilon_{2}}   \nn \eea
and the Liouville background charge $Q$ can be expressed as $Q=b+1/b$.
The conformal block in question corresponds to the factorisation of
the four-point function in which the operator
$V_\alpha(z)=\exp[2\alpha\phi(z)]$ appears as the intermediate state
in the S-channel as shown in Figure \ref{Qfig4}.
The external Liouville momenta are related
to the gauge theory mass parameters as,
\bea \alpha_{1}=\frac {Q}{2} + \frac{(\tilde{m}_{1}-\tilde{m}_{2})}
{2\hbar} & \qquad{} \qquad{} & \alpha_{2}=\frac{(\tilde{m}_{1}+\tilde{m}_{2})}
{2\hbar} \nn \\
\alpha_{4}=\frac {Q}{2} + \frac{({m}_{1}-{m}_{2})}
{2\hbar} & \qquad{} \qquad{} & \alpha_{3}=\frac{({m}_{1}+{m}_{2})}
{2\hbar} \nn \eea
and the internal momentum is given in terms of the Coulomb branch
modulus as $\alpha=Q/2+a/\hbar$. The conformal dimensions of
external and intermediate states are given by
\begin{align}
  \D_i = \alpha_i (Q- \alpha_i)\, \qquad
  \D = \alpha (Q-\alpha) \ .
\end{align}
Note that the conformal block is invariant under individual flips of
Liouville momenta $\alpha$ by $Q-\alpha$, which can be understood as
the Weyl group of $SU(2)$ gauge symmetry as well as flavour symmetries.


\paragraph{}
The Nekrasov-Shatashvili limit is one in which
$\epsilon_{2}\rightarrow 0$ with $\epsilon_{1}=\epsilon$ held
fixed. Thus $\hbar \rightarrow 0$ and $b \to \infty $, while
keeping $b\hbar = \e$ fixed. The Liouville background charge $Q\simeq b=
\epsilon/\hbar$. The Higgs branch root of Theory I is specified by the
following condition $\vec{a}=\vec{m}-\epsilon$ or,
\bea
a=m_{1}-\epsilon & \qquad{} & -a =m_{2}-\epsilon \nn
\eea
which yield $\alpha_{3}=\epsilon/\hbar$ and $\Delta_{3}=0$. Thus, in
this limit, the Higgs branch root corresponds to a special case in
which $V_{\alpha_3}(q)$ has zero dimension and therefore corresponds to
the identity operator. Now suppose we introduce a single $D2$ brane
setting $\hat{n}_{1}=1$ and $\hat{n}_{2}=0$. According to our
dictionary, $\hat{n}_{l}={n}_{l}-1$ this is dual to a point on the
Coulomb branch specified as,
\bea
a=m_{1}-2\epsilon & \qquad{} & -a =m_{2}-\epsilon \nn
\eea which yield $\alpha_{3}=3\epsilon/2\hbar$ and
$\Delta_{3}=-3\epsilon^2/4\hbar^2$. This is precisely the same
dimension as the degenerate operator,
\bea
\Phi_{2,1}(z) & = & \exp\left[-b \phi(z)\right]
\nn \eea
Interestingly, it is conjectured in \cite{AGGTV} that an insertion
$\Phi_{2,1}(z)$ is dual to the insertion of a surface operator
in the gauge theory. This is precisely consistent with our
identification of Theory II, with $\hat{n}_{1}=1$, $\hat{n}_{2}=0$
as the worldvolume theory on a single D2 brane. More general values of
$\hat{n}_{l}$ can therefore correspond to the insertion of multiple surface
operators.
\paragraph{}
The duality proposed in this paper relates the world-volume theory on a
surface operator probing the Higgs branch of a four dimensional gauge
theory with a corresponding bulk theory (ie the same four dimensional gauge
theory without surface operator on its Coulomb branch). As such it is
reminiscent of the AdS/CFT correspondence and other large-$N$
dualities. This observation can be made precise in the context of
geometric engineering where the Nekrasov partition function of
four-dimensional theory is computed by the closed topological string
on a suitable local geometry. More precisely we should consider the
closed string partition function computed using the refined
topological vertex of \cite{ref}. On the other hand, the partition
function for gauge theory in the presence of a surface operator
corresponds to an open topological string partition function
\cite{AB,Taki}. The proposed duality therefore asserts the equality of
certain open and closed topological string partition functions and it
is natural to ask if it is related to the geometric transition of
Gopakumar and Vafa \cite{GV}. Strictly speaking the latter
is defined only in the unrefined case corresponding to
$\epsilon_{1}=-\epsilon_{2}=g_{s}$ while our duality proposal applies
only to the NS limit $\epsilon_{2}\rightarrow 0$. Nevertheless there
are strong similarities which suggest that a ``refined'' geometric
transition should exist and should be equivalent in the NS limit to the duality
proposed in this paper (see also \cite{DGH}).
\begin{figure}
\centering
\includegraphics[width=140mm]{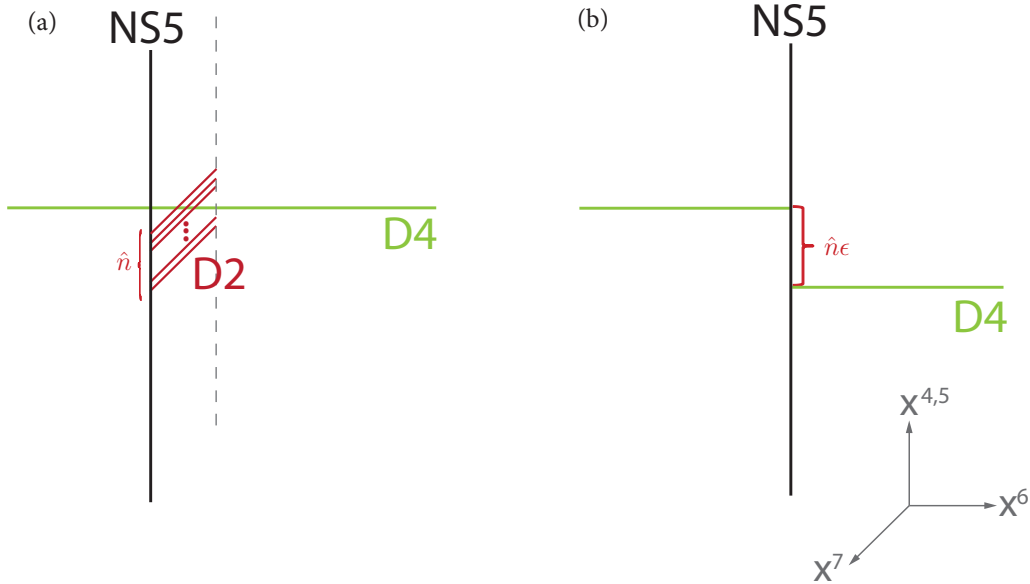}
\caption{(a) Theory II: $\hat{n}$ D2 branes suspended between a D4 and an NS5.
(b) Theory I: D4 brane breaks on NS5. }
\label{Qfig5}
\end{figure}
\paragraph{}
To understand the connection to the Geometric transition it is useful
to compare the IIA brane constructions of Theory II and Theory I shown
in Figures (\ref{Qfig3}) and (\ref{Qfig1})
respectively. Specifically we will focus on the
region of these figures corresponding to a single $U(1)$ vector
multiplet in the low-energy theory coming from a single D4 brane
and a single charged hypermultiplet which arises when the D4 brane
breaks on a single NS 5-brane. The corresponding region of Figures
(\ref{Qfig3}) and (\ref{Qfig1}) are shown in Figures
(\ref{Qfig5}(a)) and (\ref{Qfig5}(b)) respectively. This configuration is
related to the confold singularity by the standard sequence of
dualities which relate the intersecting brane and geometrical
engineering approaches to SUSY gauge theory.
In this context, the Coulomb branch of the gauge theory
corresponds to the small resolution of the conifold where an $S^{2}$
is blown up. The area of the sphere is related to the Coulomb branch
modulus $a$. The Higgs branch of the gauge theory corresponds to the
deformation of the conifold where an $S^{3}$ of non-zero size
appears. Gauge theory vortex strings are realised as D4 branes wrapped
on $S^{3}$ and extended in two dimensions of four dimensional
spacetime. Thus the duality discussed in this paper corresponds to a
transition between the deformed conifold with $\hat{n}$ wrapped branes
and the resolved conifold without branes. The resulting two sphere has
size which is linearly related to the number $\hat{n}$ of
branes. Thus the duality is of the same form as the geometric
transition. This also seems to be consistent with the relation
between the conjecture of \cite{AGGTV} and the geometric transition
suggested in \cite{DGH,Taki}.
\paragraph{}
Another interesting connection is the one between supersymmetric gauge
theory and ``holomorphic" matrix models proposed in \cite{DV1}. In
general terms, we expect
the Nekrasov partition function of Theory I to be captured by an
appropriate matrix model. Here we will make a concrete proposal for
the form of the matrix model in the NS limit. In
particular we will consider an integral over an $N\times N$ complex
matrix $\Phi$, but interpreted in the holomorphic sense of \cite{DV1},
\bea
\mathcal{Z}_{\Phi} & = & \int\,
\mathcal{D}_{\epsilon_{1},\epsilon_{2}}\Phi\,\,
\exp\left(-{\rm
    Tr}_{N}\,S(\Phi,\epsilon_{1},\epsilon_{2})\right)\ ,
\nn \eea
where the measure is $SL(N,\mathbb{C})$ invariant and can be written
in terms of the eigenvalues $\{x_{1},x_{2},\ldots,x_{N}\}$ of
$\Phi$. We propose that the leading behaviour of the resulting
integral as $\epsilon_{2}\rightarrow 0$ has the form,
\bea
\mathcal{Z}_{\Phi} & = & \int\prod_{j=1}^{N}\,dx_{j}\,
\prod_{j<k}\, \mathcal{M}\left(x_{j}-x_{k}\right)\,\,
\exp\left( -\frac{1}{\epsilon_{2}}\,
  \sum_{j=1}^{N}\mathcal{V}(x_{j})\right)\ , \nn \eea
where the measure is a deformation of the Vandermonde determinant
specified as,
\bea
\mathcal{M}(x) &  = & \exp\left(-\frac{1}{\epsilon_{2}}
\left[f(x+\epsilon)-f(x-\epsilon)\right]\right)\ ,
\nn
\eea
with $f(x)=x(\log x - 1)$ as above and potential,
\bea
\mathcal{V}(x) & = & -2\pi i \, \hat{\tau}\,x \,\,\,+\,\,\,
\sum_{l=1}^{L}\left[f(x-M_{l})-f(x-\tilde{M}_{l})\right]\ , \nn
\eea
where each of the above relations is corrected at higher orders in
$\epsilon_{2}$. On one hand we can reorganise the exponent to write
the resulting matrix integral as,
\bea
\mathcal{Z}_{\Phi} & = & \int\, \prod_{j=1}^{N}\,dx_{j}
\,\,\exp\left(\frac{1}{\epsilon_{2}}\mathcal{W}^{(II)}(x)\right)\ , \nn
\eea
where $\mathcal{W}_{II}$ is the superpotential of Theory II as given
in (\ref{spII}) which is equivalent to the Yang potential of the Bethe
Ansatz. As $\epsilon_{2}\rightarrow 0$ the matrix integral is
dominated by its saddle-points, {\it even for finite\/} $N$.
The saddle-point condition coincides
with the F-term equations of Theory II which are precisely the BAE of
the spin chain. According to the duality proposed
above the resulting saddle point value of $\log \mathcal{Z}_{\Phi}$ is
a multivalued function whose branches coincide (up to a
vacuum-independent shift) with the quantum
prepotential $\mathcal{F}(\vec{a},\epsilon)$ of Theory I evaluated at
the lattice of points $\vec{a}-\vec{m}\in \epsilon \mathbb{Z}^{L}$.
The value of the classical prepotential ${\cal F}(\vec{a})$ at any point on
the Coulomb branch can be obtained by an appropriate $N\rightarrow
\infty$ limit with $\epsilon\rightarrow 0$ and the product $N\epsilon$ fixed.
In this limit, the above matrix model is related to the proposal
of \cite{DV1} since the deformed Vandermonde determinant simplifies
and the matrix integral becomes
\bea
\mathcal{Z}_{\Phi} & = & \int \prod_{j=1}^{N}\,dx_{j}\,
\prod_{j<k}\, \left(x_{j}-x_{k}\right)^{2\beta}\,\,
\exp\left( -\frac{1}{\epsilon_{2}}\,
  \sum_{j=1}^{N}\mathcal{V}(x_{j})\right) \nn \eea
with $\beta=-\epsilon_{1}/\epsilon_{2}$. This is a
$\beta$-ensemble with a Penner-like logarithmic potential, although
involving $x\log x$ rather than $\log x$,  whose resolvent is
precisely the Seiberg-Witten 1-form,
\EQ{
L(x)dx=\sum_j\frac{dx}{x-x_j}\equiv \epsilon_1^{-1} \lambda(\e)\ .
}
It follows that in the limit $\epsilon_1\to0$, with $\epsilon_1\hat n_\ell$ fixed, we have
\EQ{
a_l-m_l=\lim_{\epsilon_1\to0}\oint_{\alpha_l}\frac{dx}{2\pi i}\,\epsilon_1\,L(x)\ .
}
The free energy in this limit is the prepotential:
\EQ{
{\cal F}(\vec a)\thicksim \lim_{\epsilon_1,\epsilon_2\to0}\epsilon_1\,\epsilon_2\log{\cal Z}_\Phi\ .
}
The resulting matrix model is precisely of the sort discussed
in \cite{DV1}. This
proposal also looks superficially similar to the matrix model reformulation of
the Nekrasov partition function derived in \cite{Sulk}, except that in
that formulation the dependence on the $\{a_l\}$ is through the
potential whereas in our matrix model it is through the filling fractions.
\paragraph{}
It is also interesting to relate the duality proposed here to an
earlier proposal for a 2d/4d duality made by two of the present
authors in \cite{DHT} (see also \cite{D1}). There the massive BPS spectrum of
the four-dimensional Theory I at the root of its Higgs branch was
related to the BPS spectrum of a $U(1)$ gauge theory in two dimensions
closely related to Theory II considered above. In contrast, in the
present case, the relation between
the superpotentials of Theories I and II takes the form,
\bea
\mathcal{W}^{(II)} & = &  \left. \mathcal{W}^{(I)}
\right|_{\vec{a}=\vec{m}-\vec{n}\epsilon}\,\,-\,\,
 \left. \mathcal{W}^{(I)}
\right|_{\vec{a}=\vec{m}-\epsilon}
\nn \eea
where as above have fixed the additive constant in the superpotential
to by subtracting the value at the root of the Higgs branch. Taking
the limit $\epsilon\rightarrow 0$ using (\ref{wa}) and (\ref{ade}) we
get,
\bea
\lim_{\epsilon\rightarrow 0}\,\,
\mathcal{W}^{(II)} & = & \left. -\sum_{l=1}^{N}
\hat{n}_{l}\vec{a}^{D}_{l}\right|_{\vec{a}=\vec{m}}
\nn \eea
Thus, in this limit,
the on-shell value of the superpotential for the two-dimensional
Theory II is related to the magnetic central charge of the
four-dimensional theory evaluated at the root of the Higgs branch. In
the abelian case where $N=\sum_{l}\hat{n}_{l}=1$, the resulting
relation has exactly the same form as that proposed in \cite{D1,DHT}.
\paragraph{}
Ideally one would also like to provide a physical explanation of the
duality along the lines of \cite{HT,Sh}. Why should the worldvolume
theory of a vortex string have any relation to the bulk theory in the
presence of the Nekrasov deformation? Here we note that, in the
presence of the specific deformation we consider, magnetic flux is
quantized in the resulting supersymmetric vacuum states even on the
Coulomb branch. Indeed, the quantum
numbers $\{n_{l}\}$ which describe the number of quanta of magnetic
flux in each low energy $U(1)$ are directly related by our conjecture
to the magnetic
fluxes $\{\hat{n}_{l}\}$ of the vortices on the Higgs branch. This
suggests the possibility of a smooth
interpolation between corresponding vacua of
Theory I and Theory II corresponding to a
variation of ${\cal N}=(2,2)$ D-term couplings,
thereby explaining the equality
of F-terms between the two theories.
\paragraph{}
Finally, we have represented the relation between the theories as a
conjecture. However, the result can be proved by formulating the BAE
as an integral equation, along the lines of the Destri-de Vega
equation \cite{Destri}. The resulting non-linear equation is then
essentially identical to the saddle-point equation that describes the
instanton partition function in the Nekrasov-Shatashvili limit
\cite{NS4d}. The recent paper \cite{Fucito:2011pn} provides a nice
description of the instanton saddle-point equation and there is a
direct link between the solution of the BAE and the instanton density
function: the pairs $(\lambda_j,\lambda^{(0)}_j)$ are precisely the
end-points of the intervals---continued into the complex plane---along
which the instanton density is non-vanishing. These issues will be
described in
detail in a companion paper \cite{us2}.
\paragraph{}
The formalism that we have developed can also be extended in various
ways: to quiver gauge theories, including the elliptic models, and to
compactified five and six dimensional ${\cal N}=2$ theories. These
generalizations will also be reported elsewhere.

\paragraph{}
\vskip 1cm
\centerline{\bf\large Acknowledgement}
\vskip 5mm
The authors acknowledge useful conversations with Stefano Bolognesi,
Heng-Yu Chen, Kazuo Hosomichi, Sergei Gukov, Misha Shifman, David Tong
and Alyosha Yung.

\newpage
\centerline{\Large \bf Appendix}

\appendix

\section{Computational Details}

\subsection{Theory I}
We present in this section the details in computation
of the on-shell twisted superpotential of Theory I,
four-dimensional $N=2$ gauge theories in $\O$-background with $\e_2=0$.
Before presenting the computation, let us first set
\begin{align}
  N = \sum_{l=1}^L (n_l-1) \ ,
\end{align}
where $N$ represents the number of D2-branes in Theory II.

\paragraph{}
One can rewrite one-loop contributions from
fundamental hyper- and vector-multiplets to the
function $\CG_\text{1-loop}$ into the following forms
\begin{align}
  \CG_\text{1-loop}^\text{fund} = & \e \sum_{l,m>l} \bigg[ \sum_{k=1-n_l}^{-1} -
  \sum_{k-1}^{n_m-1} \bigg] f\big( \frac{m_{lm}}{\e} + k \big) +
  \e \sum_l \sum_{k=1}^{n_l-1} f(-k )
  \nonumber \\ &
  +  i\pi \e \sum_{l,m>l}
  \sum_{k=1}^{n_m-1} \big( \frac{m_{lm}}{\e} + k \big)
  + \frac{LN \e}{2} \log 2\pi\ ,
\end{align}
and
\begin{align}
  \CG_\text{1-loop}^\text{vec} = - \e \sum_{l,m>l} \bigg[ \sum_{k=1+n_m-n_l}^0 +
  \sum_{k=n_m-n_l}^{-1} \bigg] f \big( \frac{m_{lm}}{\e} + k \big) + i\pi \e
  \sum_{l,m>l} \sum_{k=n_m-n_l}^{-1} \big( \frac{m_{lm}}{\e} + k \big) .
\end{align}
Summing up the above results, one can obtain
\begin{align}
  \CG_\text{1-loop}^\text{vec} + \CG_\text{1-loop}^\text{fund} = &
  \e \sum_{l,m>l} \bigg[ \sum_{k=1-n_l}^{n_m-n_l-1} - \sum_{k=1+n_m-n_l}^{n_m-1} \bigg]
  f\big( \frac{m_{lm}}{\e} + k \big) - \e \sum_l \sum_{k=1}^{n_l -1} f(k) + i\pi \e \sum_l \frac{n_l (n_l-1)}{2}
  \nonumber \\
  & + i\pi\e \sum_{l,m>l} \bigg[ \sum_{k=1}^{n_m-1}+ \sum_{k=n_m-n_l}^{-1} \bigg]
  \big( \frac{m_{lm}}{\e} + k  \big) + \frac{LN \e}{2} \log 2\pi\ ,
  \nonumber \\
  \simeq & - \e \sum_{l,m} \sum_{k=1+n_m-n_l}^{n_m-1} f\big( \frac{m_{lm}}{\e} + k \big)
  + i\pi \e \sum_{l,m>l} \bigg[ \sum_{k=1-n_l}^{n_m-n_l-1}  - \sum_{k=1}^{n_m-1}
  + \sum_{k=n_m-n_l}^{-1} \bigg] \big( \frac{m_{lm}}{\e} + k  \big)
  \nonumber \\ &
  + i\pi \e \sum_l \frac{n_l (n_l-1)}{2}  +
  \frac{LN \e}{2} \log 2\pi\ ,
\end{align}
where we used for the last equality a property of the multi-valued function $f(x)$.
One can further simplify the above expression into
\begin{align}\label{aaa}
  \CG_\text{1-loop}^\text{vec} + \CG_\text{1-loop}^\text{fund}
  = & - \e \sum_{l,m} \sum_{k=1+n_m-n_l}^{n_m-1} f\big( \frac{m_{lm}}{\e} + k \big)
  +  i\pi \sum_{l,m>l} m_{lm} (n_l-n_m ) -  i\pi \e \sum_{l,m} \sum_{k=1}^{n_m-1} k
  \nonumber \\ &
  + i\pi \e \sum_l n_l (n_l-1) + \frac{L N \e}{2} \log 2\pi
  \nonumber \\
  \simeq & - \e \sum_{l,m} \sum_{k=1+n_m-n_l}^{n_m-1} f\big( \frac{m_{lm}}{\e} + k \big)
  + i \pi L \sum_l \bigg[ (n_l -1) m_l - \frac{n_l(n_l-1)}{2} \e \bigg]
  \nonumber \\ &
  -i\pi  N \sum_l m_l + \frac{LN \e}{2} \log 2\pi\ .
\end{align}
Demanding the traceless condition of $SU(L)$,
\begin{align}
  \sum_{l=1}^L \big( m_l - n_l \e \big) = 0\ , \nonumber
\end{align}
one can replace the first term in the last line of (\ref{aaa}) as
$\sum_l m_l =  (N + L)\e$. In the $U(L)$ theory, the additional 
non-vanishing term corresponds to a constant vacuum-independent shift
of the superpotential.

\subsection{Theory II}

We now in turn discuss some computational details in
simplifying the twisted superpotential at leading order for
Theory II.
The contributions from the fundamental and adjoint chiral multiplets will be
focused in what follows.

\paragraph{}
One can massage the twisted superpotential at leading order
as follows:
\begin{align}
  \CW_\text{fund}^{(0)} = & -\e\sum_{l,m} \sum_{k=0}^{\hat n_l -1} f \big(
  \frac{M_{lm}}{\e} - k \big)
  \nonumber \\
  = & - \e \sum_{l,m>l} \bigg[
  \sum_{k=1-\hat n_l}^{0} f\big( \frac{M_{lm}}{\e} + k \big)
  + \sum_{k=0}^{\hat n_m -1} f\big( - \frac{M_{lm}}{\e} - k \big)
  \bigg] - \e \sum_{l}\sum_{k=1-\hat n_l}^{0} f\big(k\big)
  \\
  = & -\e \sum_{l,m>l} \bigg[\sum_{k=1-\hat n_l}^{0}
  - \sum_{k=0}^{\hat n_m -1}\bigg] f\big( \frac{M_{lm}}{\e} + k \big)
  - i \pi \e \sum_{l,m>l} \sum_{k=0}^{\hat n_m -1}
  \big( \frac{M_{lm}}{\e} + k \big)  - \e \sum_{l}\sum_{k=1-\hat n_l}^{0} f\big(k\big)\ .
  \nonumber
\end{align}
%
One can also show
\begin{align}
  \CW_\text{adj}^{(0)} = & \e \sum_{l,m} \sum_{s_l=1}^{\hat n_l}
  \sum_{s_m=1}^{\hat n_m} f\big( \frac{M_{lm}}{\e} -s_l + s_m -1 \big)
  \nonumber \\
  = & \e \sum_{l,m>l} \sum_{s_l,s_m} \bigg[ f\big( \frac{M_{lm}}{\e} -s_l + s_m -1 \big)
  - f \big( \frac{M_{lm}}{\e} - s_l   +s_m  + 1 \big) + i \pi \big(
  \frac{M_{lm}}{\e} - s_l   +s_m  + 1 \big) \bigg]
  \nonumber \\
  & + \e \sum_l \sum_{s_l,t_l} f \big( -s_l + t_l -1 \big)
  \nonumber \\ = &
  \e \sum_{l,m>l} \bigg[ \sum_{k=-\hat n_l}^{-1} + \sum_{k=1-\hat n_l }^0  -
  \sum_{k=\hat n_m-\hat n_l}^{\hat n_m -1} - \sum_{k=\hat n_m -\hat n_l +1}^{\hat n_m}
  \bigg] f\big( \frac{M_{lm}}{\e} + k \big)
  \nonumber \\ &
  + i\pi \e \sum_{l,m>l} \sum_{s_l,s_m} \big( \frac{M_{lm}}{\e} - s_l   +s_m  + 1 \big)
  + \e \sum_l \sum_{s_l,t_l} f\big( -s_l + t_l -1 \big)\ .
\end{align}
Here the second and the last terms in the last equality can be simplified into
\begin{align}
  i\pi \e \sum_{l,m>l} \sum_{s_l} \bigg [
  \hat n_m \big(  \frac{M_{lm}}{\e} - s_l + 1 \big)
  + \frac{\hat n_m ( \hat n_m +1 )}{2} \bigg] = & i\pi \sum_{l,m>l} \hat n_l \hat n_m M_{lm}
  + i\pi \e \sum_{l,m>l} \hat n_l \hat n_m \frac{
  2 -\hat n_l + \hat n_m }{2}\ ,
\end{align}
and
\begin{align}
  \e \sum_l &\sum_{t_l>s_l} f(t_l -s_l -1)
  - f(t_l -s_l +1 )
  + i\pi \e \sum_l \sum_{t_l>s_l}   (t_l -s_l +1)  + \e N f(-1)
  \nonumber \\
  = &\e \sum_{l} \sum_{s_l} \bigg[ f (0) + f(1)- f(n_l-s_l)
  - f(n_l-s_l +1) \bigg]  + \frac{i\pi \e}{6} \sum_l \hat n_l
  (\hat n_l - 1) (\hat n_l +4)
  \nonumber \\
  = & \e N \big( f(1) + f(-1) \big)
  - \e \sum_{l} \bigg[ \sum_{k=0}^{\hat n_l -1} + \sum_{k=1}^{\hat n_l} \bigg]
  f(k)  + \frac{i\pi \e}{6} \sum_l \hat n_l
  (\hat n_l - 1) (\hat n_l +4) \ .
\end{align}
It therefore implies that
\begin{align}
  \CW_\text{adj}^{(0)} = &
  \e \sum_{l,m>l} \bigg[ \sum_{k=-\hat n_l}^{-1} + \sum_{k=1-\hat n_l }^0  -
  \sum_{k=\hat n_m-\hat n_l}^{\hat n_m -1} - \sum_{k=\hat n_m -\hat n_l +1}^{\hat n_m}
  \bigg] f\big( \frac{M_{lm}}{\e} + k \big)
  \nonumber \\ &
  + i\pi \sum_{l,m>l} \hat n_l \hat n_m M_{lm}
  + i\pi \e \sum_{l,m>l} \hat n_l \hat n_m \frac{
  2 -\hat n_l + \hat n_m }{2}
  \nonumber \\ &
  + i\pi \e N  - \e\sum_l
  \bigg[ \sum_{k=0}^{\hat n_l -1} + \sum_{k=1}^{\hat n_l} \bigg] f(k)
  + \frac{i\pi \e}{6} \sum_l \hat n_l
  (\hat n_l - 1) (\hat n_l +4) \ .
\end{align}
Collecting all the results, one can obtain
\begin{align}
  \CW_\text{fund}^{(0)} + \CW_\text{adj}^{(0)} &
  =  \e \sum_{l,m>l} \bigg[ \sum_{k=-\hat n_l}^{
  \hat n_m -\hat n_l -1} - \sum_{k=\hat n_m -\hat n_l+1}^{\hat n_m} \bigg]
  f\big(\frac{M_{lm}}{\e} + k \big) - i\pi\e\sum_{l,m>l} \sum_{k=0}^{\hat n_m-1}
  \big( \frac{M_{lm}}{\e} + k \big)
  \nonumber \\ &
  + i\pi \sum_{l,m>l} \hat n_l \hat n_m M_{lm}
  + i\pi \e \sum_{l,m>l} \hat n_l \hat n_m \frac{
  2 -\hat n_l + \hat n_m }{2}
  \nonumber \\ &
  + i\pi \e N - i\pi \e \sum_l \frac{\hat n_l (\hat n_l -1)}{2}
  + \frac{i\pi \e}{6} \sum_l \hat n_l (\hat n_l - 1) (\hat n_l +4)
  - \e \sum_l \sum_{k=1}^{\hat n_l} f(k)
  \nonumber \\ = &
  - \e \sum_{l,m} \sum_{k=\hat n_m -\hat n_l +1}^{\hat n_m}
  f\big( \frac{M_{lm}}{\e} + k \big)
  + i\pi \sum_{l,m>l} \hat n_l \hat n_m M_{lm}
  + i\pi \e \sum_{l,m>l} \hat n_l \hat n_m \frac{
  \hat n_m -\hat n_l  }{2}
  \nonumber \\ &
  + \frac{i\pi \e}{6} \sum_l \hat n_l (\hat n_l - 1) (\hat n_l +1)
  + i\pi \e N \ .
\end{align}
It can be further simplified into
\begin{align}
  \CW_\text{fund}^{(0)} + \CW_\text{adj}^{(0)} &  =
  - \e \sum_{l, m} \sum_{k=\hat n_m -\hat n_l +1}^{\hat n_m}
  f\big( \frac{M_{lm}}{\e} + k \big)
  + i\pi \big( N + 1\big) \sum_l \sum_{s_l}  \big( M_l - (s_l-1)\e \big)
  \nonumber \\ &
  - i\pi \sum_l \big( \hat n_l +1 \big) \hat n_l M_l
  + \frac{2i\pi\e}{3} \sum_l  \hat n_l (\hat n_l - 1) (\hat n_l +1)
  \nonumber \\ &
  - 2\pi i \sum_l \sum_{s_l} \big( M_l - (s_l-1)\e \big) \sum_{m<l} \hat n_m
  + i\pi \e N
  \\ \simeq &
  - \e \sum_{l, m} \sum_{k=\hat n_m -\hat n_l +1}^{\hat n_m}
  f\big( \frac{M_{lm}}{\e} + k \big)
  + i\pi \big( N + 1\big) \sum_l \sum_{s_l}  \big( M_l - (s_l-1)\e \big)
  + i\pi \e N\nonumber
\end{align}
Again, we used for the last equality a property of multi-valued function $f(x)$.

%
%
%
%
%


\begin{thebibliography}{99}

\bibitem{WM}
  E.Witten,
  ``{\it Solutions of Four-Dimensional Field Theories via M Theory},''
  Nucl.\ Phys.\  {\bf B500 } (1997)  3-42.
  [hep-th/9703166].

\bibitem{Wphase} E.~Witten,
  ``{\it Phases of N=2 Theories in Two-Dimensions},''
  Nucl.\ Phys.\  {\bf B403 } (1993)  159-222.
  [hep-th/9301042].

\bibitem{NS2d}
  N.A.~Nekrasov and S.L.~Shatashvili,
  ``{\it Supersymmetric Vacua and Bethe Ansatz},''
  Nucl.\ Phys.\ Proc.\ Suppl.\  {\bf 192-193} (2009) 91
  [arXiv:0901.4744 [hep-th]]. \vspace*{0.2cm}\\
  N.A.~Nekrasov and S.L.~Shatashvili,
  ``{\it Quantum Integrability and Supersymmetric Vacua},''
  Prog.\ Theor.\ Phys.\ Suppl.\  {\bf 177} (2009) 105
  [arXiv:0901.4748 [hep-th]].



\bibitem{NS4d}
  N.A.~Nekrasov and S.L.~Shatashvili,
  ``{\it Quantization of Integrable Systems and Four Dimensional
Gauge Theories},''
  arXiv:0908.4052 [hep-th].

\bibitem{M3}
  A.~Marshakov, A.~Mironov, A.~Morozov,
  ``{\it On AGT Relations with Surface Operator Insertion and Stationary
Limit of Beta-Ensembles},''
  Teor.\ Mat.\ Fiz.\  {\bf 164 } (2010)  3-27.
  [arXiv:1011.4491 [hep-th]].



\bibitem{Gors}
  A.~Gorsky, I.~Krichever, A.~Marshakov, A.~Mironov, A.~Morozov,
  ``{\it Integrability and Seiberg-Witten Exact Solution},''
  Phys.\ Lett.\  {\bf B355 } (1995)  466-474.
  [hep-th/9505035].

\bibitem{Mart}  E.J.~Martinec, N.P.~Warner,
  ``{\it Integrable Systems and Supersymmetric Gauge Theory},''
  Nucl.\ Phys.\  {\bf B459 } (1996)  97-112.
  [hep-th/9509161]. \vspace*{0.2cm}\\
  E.J.~Martinec,
  ``{\it Integrable Structures in Supersymmetric Gauge and String Theory},''
  Phys.\ Lett.\  {\bf B367 } (1996)  91-96.
  [hep-th/9510204].

\bibitem{DW}
  R.~Donagi, E.~Witten,
  ``{\it Supersymmetric Yang-Mills Theory and Integrable Systems},''
  Nucl.\ Phys.\  {\bf B460 } (1996)  299-334.
  [hep-th/9510101].

\bibitem{Marsh1} A.~Gorsky, A.~Marshakov, A.~Mironov, A.~Morozov,
  ``{\it N=2 Supersymmetric QCD and Integrable Spin Chains: Rational Case
  $N(f) < 2N(c)$,}''
  Phys.\ Lett.\  {\bf B380 } (1996)  75-80.
  [hep-th/9603140]. \vspace*{0.2cm}\\
  A.~Gorsky, S.~Gukov, A.~Mironov,
  ``{\it Multiscale N=2 SUSY Field Theories, Integrable Systems
  and Their Stringy/Brane Origin. 1.},''
  Nucl.\ Phys.\  {\bf B517 } (1998)
  409-461. [hep-th/9707120]. \vspace*{0.2cm}\\
  A.~Gorsky, S.~Gukov, A.~Mironov,
  ``{\it SUSY Field Theories, Integrable Systems and Their
  Stringy/Brane Origin. 2.},''
  Nucl.\ Phys.\  {\bf B518 } (1998)  689-713.
  [hep-th/9710239].

\bibitem{DV}
  R.~Dijkgraaf and C.~Vafa,
  ``{\it Matrix Models, Topological Strings, and Supersymmetric 
  Gauge Theories},''
  Nucl.\ Phys.\  B {\bf 644} (2002) 3
  [arXiv:hep-th/0206255]. \vspace*{0.2cm}\\
  R.~Dijkgraaf and C.~Vafa,
  ``{\it On Geometry and Matrix Models},''
  Nucl.\ Phys.\  B {\bf 644} (2002) 21
  [arXiv:hep-th/0207106]. \vspace*{0.2cm}\\
  R.~Dijkgraaf and C.~Vafa,
  ``{\it A Perturbative Window into Non-Perturbative Physics},''
  arXiv:hep-th/0208048.

\bibitem{Sh}
  M.~Shifman, A.~Yung,
  ``{\it NonAbelian string junctions as confined monopoles},''
  Phys.\ Rev.\  {\bf D70 } (2004)  045004.
  [hep-th/0403149].

\bibitem{HT}
  A.~Hanany, D.~Tong,
  ``{\it Vortex Strings and Four-Dimensional Gauge Dynamics},''
  JHEP {\bf 0404 } (2004)  066.
  [hep-th/0403158].

\bibitem{HT1} A.~Hanany, D.~Tong,
  ``{\it Vortices, Instantons and Branes},''
  JHEP {\bf 0307 } (2003)  037.
  [hep-th/0306150].


\bibitem{RevVS} D. Tong, ''{\it TASI Lectures on Solitons},''
 arXiv:hep-th/0509216. \\
M. Shifman and A. Yung,
''{\it Supersymmetric Solitons},''
Rev.\ Mod.\ Phys. {\bf 79} 1139 (2007)
[arXiv:hep-th/0703267] 


\bibitem{SY3} M. Shifman and A. Yung,
''{\it Non-Abelian semilocal strings in N=2 supersymmetric QCD},''
Phys. Rev. D73 (2006) 125012 [hep-th/0603134].

\bibitem{SVY} M.~Shifman, W.~Vinci, A.~Yung, 
``{\it Effective World-Sheet Theory for Non-Abelian Semilocal Strings in 
N = 2 Supersymmetric QCD},''
  Phys.\ Rev.\  {\bf D83 } (2011)  125017.
  [arXiv:1104.2077 [hep-th]].

\bibitem{HoriTong} K.~Hori, D.~Tong,
 ``{\it 
Aspects of Non-Abelian Gauge Dynamics in Two-Dimensional N=(2,2) Theories},''
  JHEP {\bf 0705 } (2007)  079.
  [hep-th/0609032].


\bibitem{Ext}
  A.~Marshakov, N.~Nekrasov,
  ``{\it Extended Seiberg-Witten Theory and Integrable Hierarchy},''
  JHEP {\bf 0701 } (2007)  104.
  [hep-th/0612019].

\bibitem{Col}
  S.~R.~Coleman,
  ``{\it More About the Massive Schwinger Model},''
  Annals Phys.\  {\bf 101 } (1976)  239.


\bibitem{SW2}
  N.~Seiberg, E.~Witten,
  ``{\it Monopoles, Duality and Chiral Symmetry Breaking in N=2
  Supersymmetric QCD},''
  Nucl.\ Phys.\  {\bf B431 } (1994)  484-550.
  [hep-th/9408099].

\bibitem{N1}
  N.~A.~Nekrasov,
  ``{\it Seiberg-Witten Prepotential from Instanton Counting},''
  Adv.\ Theor.\ Math.\ Phys.\  {\bf 7 } (2004)  831-864.
  [hep-th/0206161].


\bibitem{NO}
  N.~Nekrasov and A.~Okounkov,
  ``{\it Seiberg-Witten Theory and Random Partitions},''
  arXiv:hep-th/0306238.

\bibitem{Faddeev}
  L.D.~Faddeev,
  ``{\it How Algebraic Bethe Ansatz Works for Integrable Model},''
  [hep-th/9605187].

\bibitem{Derk}
  S.~E.~Derkachov, G.~P.~Korchemsky and A.~N.~Manashov,
  ``{\it Separation of Variables for the Quantum SL(2,R) Spin Chain,}''
  JHEP {\bf 0307} (2003) 047
  [arXiv:hep-th/0210216].

\bibitem{FadKor}
  L.~D.~Faddeev and G.~P.~Korchemsky,
  ``{\it High-energy QCD as a Completely Integrable Model},''
  Phys.\ Lett.\  B {\bf 342} (1995) 311
  [arXiv:hep-th/9404173].

\bibitem{AGT}
  L.F.~Alday, D.~Gaiotto and Y.~Tachikawa,
  ``{\it Liouville Correlation Functions from
Four-dimensional Gauge Theories},''
  Lett.\ Math.\ Phys.\  {\bf 91} (2010) 167
  [arXiv:0906.3219 [hep-th]].

\bibitem{AGGTV}
  L.F.~Alday, D.~Gaiotto, S.~Gukov, Y.~Tachikawa, H.~Verlinde,
  ``{\it Loop and Surface Operators in N=2 Gauge Theory and Liouville
  Modular Geometry},''
  JHEP {\bf 1001}, 113 (2010).

\bibitem{DGH}
  T.~Dimofte, S.~Gukov, L.~Hollands,
  ``{\it Vortex Counting and Lagrangian 3-manifolds},''
  [arXiv:1006.0977 [hep-th]].

\bibitem{Yoshida}
  Y.~Yoshida,
  ``{\it Localization of Vortex Partition Functions in $\mathcal{N}=(2,2)$
Super Yang-Mills theory},''
[arXiv:1101.0872 [hep-th]].

\bibitem{Bonelli}
  G.~Bonelli, A.~Tanzini, J.~Zhao,
  ``{\it Vertices, Vortices and Interacting Surface Operators},''

  [arXiv:1102.0184 [hep-th]].


\bibitem{GV}
  R.~Gopakumar, C.~Vafa,
  ``{\it On the Gauge Theory / Geometry Correspondence},''
  Adv.\ Theor.\ Math.\ Phys.\  {\bf 3 } (1999)  1415-1443.
  [hep-th/9811131].

\bibitem{DV1} R.~Dijkgraaf, C.~Vafa,
  ``{\it Toda Theories, Matrix Models, Topological Strings, and N=2 Gauge
  Systems},'' [arXiv:0909.2453 [hep-th]].


\bibitem{D1} N.~Dorey,
  ``{\it The BPS Spectra of Two-Dimensional Supersymmetric
  Gauge Theories with Twisted Mass Terms},''
  JHEP {\bf 9811 } (1998)  005.
  [hep-th/9806056].

\bibitem{DHT}
  N.~Dorey, T.J.~Hollowood, D.~Tong,
  ``{\it The BPS Spectra of Gauge Theories in Two-Dimensions and
Four-Dimensions},''
  JHEP {\bf 9905 } (1999)  006.
  [hep-th/9902134].

\bibitem{Lee:2009fc}
  S.~Lee, P.~Yi,
  ``{\it A Study of Wall-Crossing: Flavored Kinks in D=2 QED},''
  JHEP {\bf 1003}, 055 (2010).
  [arXiv:0911.4726 [hep-th]].

\bibitem{BGK1}
  A.V.~Belitsky, A.S.~Gorsky and G.P.~Korchemsky,
  ``{\it Gauge / String Duality for QCD Conformal Operators},''
  Nucl.\ Phys.\  B {\bf 667} (2003) 3
  [arXiv:hep-th/0304028].

\bibitem{BBGK}
  A.V.~Belitsky, V.M.~Braun, A.S.~Gorsky and G.P.~Korchemsky,
  ``{\it Integrability in QCD and Beyond},''
  Int.\ J.\ Mod.\ Phys.\  A {\bf 19}, 4715 (2004)
  [arXiv:hep-th/0407232].

\bibitem{BGK2}
  A.V.~Belitsky, A.S.~Gorsky and G.P.~Korchemsky,
  ``{\it Logarithmic Scaling in Gauge / String Correspondence},''
  Nucl.\ Phys.\  B {\bf 748} (2006) 24
  [arXiv:hep-th/0601112].

\bibitem{Marsh}
  A.~Mironov, A.~Morozov,
  ``{\it Nekrasov Functions and Exact Bohr-Zommerfeld Integrals},''
  JHEP {\bf 1004 } (2010)  040.
  [arXiv:0910.5670 [hep-th]]. \vspace*{0.2cm}\\
  A.~Mironov, A.~Morozov,
  ``{\it Nekrasov Functions from Exact BS Periods: The Case of $SU(N)$},''
  J.\ Phys.\ A {\bf A43 } (2010)  195401.
  [arXiv:0911.2396 [hep-th]]. \vspace*{0.2cm}\\
  A.~Mironov, A.~Morozov, S. Shakirov,
  ``{\it Matrix Model Conjecture for
Exact BS Periods and Nekrasov Functions},''
  JHEP {\bf 1002 } (2010)  030.
  [arXiv:0911.5721 [hep-th]].

\bibitem{DKM}
  N.~Dorey, V.~V.~Khoze, M.~P.~Mattis,
  ``On N=2 supersymmetric QCD with four flavors,''
  Nucl.\ Phys.\  {\bf B492 } (1997)  607-622.
  [hep-th/9611016].


\bibitem{Pop}
  A.~Popolitov,
  ``{\it On Relation between Nekrasov Functions and
BS Periods in Pure SU(N) Case},''
 [arXiv:1001.1407 [hep-th]].

\bibitem{Zenk}
  Y.~Zenkevich,
  ``{\it Nekrasov Prepotential with Fundamental Matter from the
Quantum Spin Chain},''
[arXiv:1103.4843 [math-ph]].

\bibitem{Taki1}
  K.~Maruyoshi, M.~Taki,
  ``{\it Deformed Prepotential, Quantum Integrable System and
Liouville Field Theory},''
  Nucl.\ Phys.\  {\bf B841 } (2010)  388-425.
  [arXiv:1006.4505 [hep-th]].

\bibitem{Sulk}
  P.~Sulkowski,
  ``{\it Matrix models for Beta-Ensembles from Nekrasov Partition Functions},''
  JHEP {\bf 1004 } (2010)  063.
  [arXiv:0912.5476 [hep-th]].

\bibitem{ref}
  A.~Iqbal, C.~Kozcaz, C.~Vafa,
  ``{\it The Refined Topological Vertex},''
  JHEP {\bf 0910 } (2009)  069.
  [hep-th/0701156].

\bibitem{AB}
  C.~Kozcaz, S.~Pasquetti, N.~Wyllard,
  ``{\it A \& B Model Approaches to Surface Operators and Toda Theories},''
  JHEP {\bf 1008 } (2010)  042.
  [arXiv:1004.2025 [hep-th]].

\bibitem{Taki}
  M.~Taki,
  ``{\it Surface Operator, Bubbling Calabi-Yau and AGT Relation},''
  [arXiv:1007.2524 [hep-th]].

\bibitem{Hanany:1997vm}
  A.~Hanany and K.~Hori,
  ``{\it Branes and N = 2 Theories in Two Dimensions},''
  Nucl.\ Phys.\  B {\bf 513}, 119 (1998)
  [arXiv:hep-th/9707192].

\bibitem{Destri}
  C.~Destri and H.J.~De Vega,
  ``{\it Unified Approach To Thermodynamic Bethe Ansatz And
Finite Size Corrections
  For Lattice Models And Field Theories},''
  Nucl.\ Phys.\  B {\bf 438}, 413 (1995)
  [arXiv:hep-th/9407117].

\bibitem{Pog} R.~Poghossian,
  ``{\it Deforming SW curve},''
  [arXiv:1006.4822 [hep-th]].
 
\bibitem{Fucito:2011pn}
  F.~Fucito, J.~F.~Morales, R.~Poghossian and D.~R.~Pacifici,
  ``{\it Gauge theories on $\Omega$-backgrounds from non commutative
Seiberg-Witten
  curves\/},''
  arXiv:1103.4495 [hep-th].

\bibitem{us2}
H.~Y.~Chen, N. Dorey, T. Hollowood and S. Lee, {\it to appear\/}.

\end{thebibliography}
\end{document}